\documentclass[twocolumn,floatfix,a4paper,showpacs,showkeys,nofootinbib,reprint]{revtex4-1}

\usepackage{epsfig}
\usepackage{latexsym}
\usepackage{xspace}
\usepackage[colorlinks=true,linktocpage=true,linkcolor=blue,citecolor=blue,allcolors=blue]{hyperref}
\usepackage[utf8]{inputenc}
\usepackage{indentfirst}
\usepackage{enumerate}
\usepackage{color}
\usepackage{todonotes}

\usepackage[caption=false,position=top]{subfig}

\usepackage{amsmath}
\usepackage{amssymb}
\usepackage[english]{babel}
\usepackage{url}

\newcommand{\bs}{\boldsymbol}

\newcommand{\bvar}{\mathbf}

\newcommand{\eq}[1]{\begin{align} #1 \end{align}}
\newcommand{\mean}[1]{\langle #1 \rangle}

\newcommand{\ktwonet}[1]{\kappa_2[#1 - \bar{#1}]/\mean{#1 + \bar{#1}}}

\begin{document}


\title{
Particlization of an interacting hadron resonance gas
with global conservation laws\\
for event-by-event fluctuations in heavy-ion collisions
}

\author{Volodymyr Vovchenko}
\affiliation{Nuclear Science Division, Lawrence Berkeley National Laboratory, 1 Cyclotron Road, Berkeley, CA 94720, USA}

\author{Volker Koch}
\affiliation{Nuclear Science Division, Lawrence Berkeley National Laboratory, 1 Cyclotron Road, Berkeley, CA 94720, USA}

\begin{abstract}
We revisit the problem of particlization of a QCD fluid into hadrons and resonances at the end of the fluid dynamical stage in relativistic heavy-ion collisions in a context of fluctuation measurements.
The existing methods sample an ideal hadron resonance gas, therefore, they do not capture the non-Poissonian nature of the grand-canonical fluctuations, expected due to QCD dynamics such as the chiral transition or QCD critical point.
We address the issue by partitioning the particlization hypersurface into locally grand-canonical fireballs populating the space-time rapidity axis that are constrained by global conservation laws.
The procedure allows to quantify the effect of global conservation laws, volume fluctuations, thermal smearing and resonance decays on fluctuation measurements in various rapidity acceptances, and can be used in fluid dynamical simulations of heavy-ion collisions.
As a first application, we study event-by-event fluctuations in heavy-ion collisions at the LHC
using an excluded volume hadron resonance gas model matched to lattice QCD susceptibilities, with a focus on (pseudo)rapidity acceptance dependence of net baryon, net proton, and net charge cumulants.
We point out large differences between net proton and net baryon cumulant ratios that make direct comparisons between the two unjustified.
We observe that the existing experimental data on net-charge fluctuations at the LHC shows a strong suppression relative to a hadronic description.
\end{abstract}



\maketitle


\section{Introduction}

Event-by-event fluctuations in relativistic heavy-ion collisions have long been considered sensitive experimental probes of the QCD phase structure~\cite{Stephanov:1998dy,Stephanov:1999zu,Jeon:2000wg,Asakawa:2000wh}.
At the highest collision energies achievable at the LHC and RHIC they can be used to analyze the QCD chiral crossover transition at small baryon densities~\cite{Friman:2011pf}.
The equilibrium fluctuations of the QCD conserved charges in the grand-canonical ensemble have been computed at $\mu_B = 0$ from first principles, via lattice gauge theory simulations~\cite{Borsanyi:2011sw,Bazavov:2012jq}.
An appropriately performed comparison between experimental measurements and lattice QCD predictions can, in principle, establish whether a locally equilibrated QCD matter is indeed created in experiment.
At lower collision energies, the fluctuations are used in the experimental search for the hypothetical QCD critical point and the first-order phase transition at finite baryon density.
This is motivated by the fact that fluctuations, in particular the net proton cumulants of higher order, are increasingly sensitive to the proximity of the critical point~\cite{Hatta:2003wn,Stephanov:2008qz}. 
The corresponding measurements are in the focus of several experimental programs, including beam energy scans performed at RHIC~\cite{Bzdak:2019pkr,Adam:2020unf} and CERN-SPS~\cite{Gazdzicki:2015ska}. 
The experimental data in the literature includes second order cumulants, both diagonal~\cite{Alt:2007jq,Adamczyk:2017wsl,Acharya:2019izy,Adam:2020kzk} and off-diagonal~\cite{Anticic:2013htn,Anticic:2015fla,Adam:2019xmk}, as well as higher-order fluctuation measures~\cite{Adamczyk:2013dal,Adamczyk:2014fia,Adam:2020unf,Adamczewski-Musch:2020slf}.

A proper theoretical modeling is crucial for interpreting the experimental data.
It is not uncommon in the literature to directly compare the theoretical fluctuations evaluated in the grand-canonical ensemble with experimental measurements~\cite{Karsch:2010ck,Bazavov:2012vg,Borsanyi:2014ewa,Alba:2014eba,Fukushima:2014lfa,Albright:2015uua,Fu:2016tey,Almasi:2017bhq,Vovchenko:2017ayq,Bellwied:2019pxh}.
Such comparisons, however, have several important drawbacks.
For one thing, the experimental measurements are performed in momentum space whereas the theoretical approaches operate in configuration space.
Cuts in the momentum space may be identified with the coordinate space if strong space-momentum correlations are present, for instance due to Bjorken flow, but even in this case a degree of smearing will be present because of the thermal motion~\cite{Ling:2015yau,Ohnishi:2016bdf}.
Event-by-event fluctuations, especially the high-order cumulants, are strongly affected by global conservation laws~\cite{Bleicher:2000ek,Begun:2006uu,Bzdak:2012an}, requiring large corrections to the grand-canonical distributions.
Other mechanisms include volume fluctuations~\cite{Gorenstein:2011vq,Skokov:2012ds,Braun-Munzinger:2016yjz}, finite system size~\cite{Poberezhnyuk:2020ayn}, 
as well as non-equilibrium dynamics such as memory effects~\cite{Mukherjee:2015swa} or hadronic phase evolution~\cite{Steinheimer:2016cir}.
Proper modeling of these effects is thus required for analyzing the experimental data quantitatively.

The standard approach to describe the evolution of strongly interacting QCD matter created in heavy-ion collisions is relativistic fluid dynamics~\cite{Gale:2013da,Romatschke:2017ejr}.
The hydrodynamic description terminates at a so-called particlization stage~\cite{Huovinen:2012is}, where the  QCD fluid is transformed into an expanding gas of hadrons and resonances.
This picture forms the basis of the hybrid models of heavy-ion collisions~\cite{Petersen:2008dd,Song:2010aq} and it works quite well in describing the spectra and flow of bulk hadrons measured in a broad range of collision energies~\cite{Schenke:2010nt,Song:2011hk,Shen:2011eg,Karpenko:2012yf}.

Event-by-event fluctuations of hadron yields, on the other hand, are seldom analyzed in the hydro picture.
The yields of hadrons and resonances are usually sampled in each fluid element from a Poisson distribution.
Because the Poisson distribution is additive, this means that the yields of all hadron species in the full space follow the Poisson distribution as well.
This picture corresponds to the multiplicity distribution of an ideal Maxwell-Boltzmann hadron resonance gas~(HRG) in the grand-canonical ensemble.
Most hydro simulations use this type of sampling~\cite{Kisiel:2005hn,Shen:2014vra,Karpenko:2015xea,Bernhard:2018hnz}.
More advanced procedures incorporate exact conservation of the QCD conserved charges and/or energy-momentum~\cite{Becattini:2003ft,Becattini:2004rq,Schwarz:2017bdg,Oliinychenko:2019zfk,Oliinychenko:2020cmr}, however, these procedures are still restricted to the equation of state of an ideal HRG.
The existing methods, therefore, are not suitable to analyze the fluctuation signals of any effect that goes beyond the physics of an ideal hadron gas.

Interacting HRG models, on the other hand, offer more flexibility.
For instance, an HRG model with excluded volume corrections can describe the lattice QCD cumulants of net baryon distribution in vicinity of the chemical freeze-out at $\mu_B = 0$~\cite{Vovchenko:2017xad,Vovchenko:2017drx}, which the ideal HRG model cannot.
Another example is HRG model with van der Waals interactions, which captures the physics of nuclear liquid-gas transition at large $\mu_B$~\cite{Vovchenko:2016rkn,Vovchenko:2017ayq}.
It is the purpose of this work to formulate a particlization routine appropriate to describe event-by-event fluctuations encoded in the equation of state of such interacting HRG models.

The paper is organized as follows.
In Sec.~\ref{sec:sampler} we introduce a method for sampling an interacting HRG at particlization stage of heavy-ion collisions that we call subensemble sampler.
Sec.~\ref{sec:evhrg} describes the technical details of sampling an excluded volume HRG model that we study this work as an example.
In Sec.~\ref{sec:LHC} the subensemble sampler is applied for the description of net baryon and net proton fluctuations in heavy-ion collisions at LHC energied.
Discussion and summary in \ref{sec:summary} close the article.

\section{Subensemble sampler}
\label{sec:sampler}

Consider the particlization stage of heavy-ion collisions at the end of the ideal hydrodynamic evolution.
This stage is characterized by a hypersurface $\sigma(x)$, where the space-time coordinate $x$ is taken in the Milne basis, $x = (\tau, r_x, r_y, \eta_s)$.
Here $\tau = \sqrt{t^2 -r_z^2}$ and $\eta_s = \frac{1}{2} \ln \frac{t+r_z}{t-r_z}$ are the longitudinal proper time and space-time rapidity, respectively, $r_x$, $r_y$, and $r_z$ are the Cartesian coordinates.
The QCD matter is assumed in local thermodynamic equilibrium at each point $x$ on this hypersurface.\footnote{In a more general case the deviations from local equilibrium are described using viscous corrections.} 
As the fluid is converted into hadrons at this stage, the equation of state is described by hadron and resonance degrees of freedom, i.e. this has to be a variant of the hadron resonance gas model matched to the actual QCD equation of state at each point on the hypersurface. 

Let us denote $\mathcal{Z}^{\rm HRG} (T,V,\bs \mu)$ as the grand partition function of a hadron resonance gas at temperature $T$, volume $V$, and chemical potentials $\bs \mu = (\mu_B,\mu_Q,\mu_S)$, and $P^{\rm hrg}(\{N_i\}_{i=1}^f; T,V,\bs \mu)$ as the corresponding multiplicity distribution for all hadron species.
Here $f$ is the number of different hadron species.
In case of the commonly used ideal HRG model $P^{\rm hrg}$ has a form of a multi-Poisson distribution where the Poisson means correspond to the mean multiplicities of primordial hadrons and resonances.
Most particlization routines work with the multi-Poisson distribution of the ideal HRG model.
However, $P^{\rm hrg}$ will differ from the multi-Poisson distribution in a more general case of a non-ideal HRG.
Thus, in the present work we generalize the particlization routine for arbitrary hadron multiplicity distributions.

\subsection{Uniform fireball}

Let us first consider a case of the grand-canonical ensemble, where the global conservations laws are enforced on average.
Later on we will relax this assumption to incorporate exact global conservation.

If we further assume for the time being that the intensive thermal parameters $T$, $\mu_B$, $\mu_Q$, and $\mu_S$ are the same across the entire fireball, and the partition function of the entire system coincides with the grand partition function $\mathcal{Z}^{\rm HRG}$ of a uniform HRG:
\eq{
Z_{\rm tot}^{\rm gce, unif} = \mathcal{Z}^{\rm HRG} (T,V,\bs \mu).
}
Here 
\eq{
\mathcal{Z}^{\rm HRG} (T,V,\bs \mu) = \sum_{\bvar Q} \, e^{\bs  \mu \cdot \bvar Q} \, Z^{\rm HRG} (T,V,\bvar Q)
}
with $Z^{\rm HRG} (T,V,\bvar Q)$ being the canonical partition function of the HRG model with $\bvar Q = (B,Q,S)$,
and
\eq{
V = \int d \sigma_\mu(x) \, u^{\mu} (x)
}
is the effective system volume at particlization.

The single-particle momentum distribution function is given by the Cooper-Frye formula~\cite{Cooper:1974mv}:
\eq{\label{eq:CF}
E_p \, \frac{dN_i}{d^3p} = \int d \sigma_\mu (x) \, p^\mu \, f_i(x, p)~.
}
Here $f_i(x, p)$ is the single-particle distribution function.
In the following we neglect quantum statistics and viscous corrections but take into account the possibility of interactions between hadrons.
We assume that the distribution function takes the following general form\footnote{Here we neglect the possible modifications of the momentum distribution due to interactions.}
\eq{
f_i(x, p) = \frac{\lambda_i^{\rm int}(T,\bs \mu)}{(2\pi)^3} \, \exp\left(\frac{\mu_i - p^\mu  u_\mu(x)}{T} \right).
}
Here $\mu_i = b_i \mu_B + q_i \mu_Q + s_i \mu_S$, $u_\mu(x)$ is the flow velocity profile, and $\lambda_i^{\rm int}(T,\bs \mu)$ is a correction factor which describes deviations from the ideal gas distribution function induced by interactions.
The explicit form of this factor depends on the interacting HRG model under consideration.
The mean particle number $\mean{N_i}$ is obtained by integrating Eq.~\eqref{eq:CF} over the momenta:
\eq{
\mean{N_i} & = \lambda_i^{\rm int}(T,\bs \mu) \, \frac{d_i m_i^2 T}{2\pi^2} \, K_2(m_i/T) \, e^{\mu_i/T} \, V, \\
& = \lambda_i^{\rm int}(T,\bs \mu) \, \mean{N_i}^{\rm id}.
}

The full space hadron multiplicity distribution is given by the multiplicity distribution of the grand-canonical HRG:
\eq{
P^{\rm gce, unif}(\{N_i\}_{i=1}^f) = P^{\rm hrg}(\{N_i\}_{i=1}^f; T,V,\bs \mu).
}

\subsection{Partition in rapidities}

\begin{figure}[t]
  \centering
  \includegraphics[width=.49\textwidth]{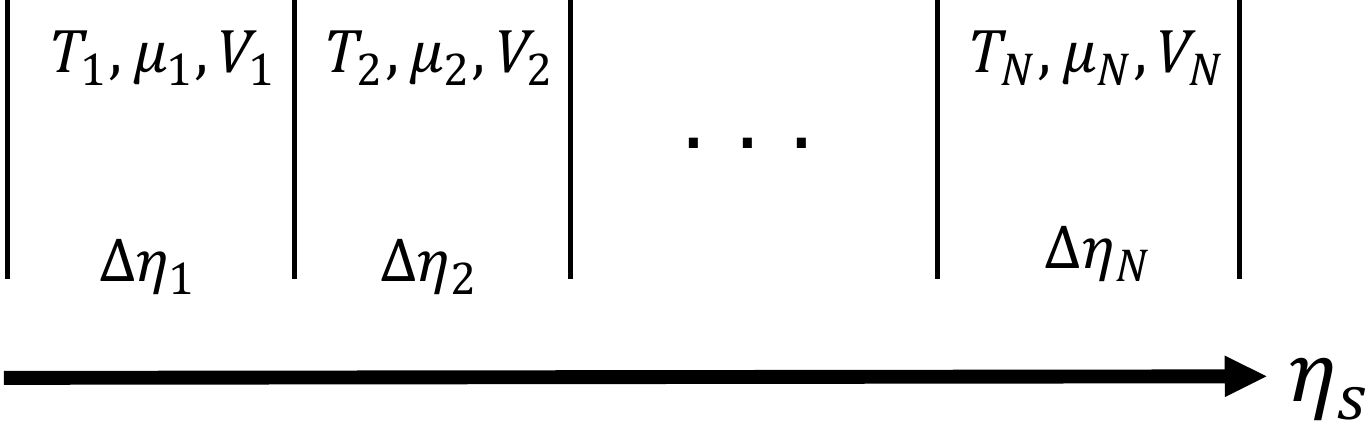}
  \caption{
   A schematic view of the partition of the space-time rapidity axis at particlization into $N$ locally grand-canonical subvolumes, each characterized by values of the local temperature $T_j$, the chemical potential $\mu_j$, and the volume $V_j$.
  }
  \label{fig:partition}
\end{figure}

Let us now split the hypersurface into $s$ slices along the space-time rapidity axis~(see Fig.~\ref{fig:partition}).
The boundaries of each slices are $\eta_{j}^{\rm min}$ and $\eta_{j}^{\rm max} > \eta_{j}^{\rm min}$.
Furthermore, one has $\eta_{j}^{\rm min} = \eta_{j-1}^{\rm max}$ for $j > 1$, and $\eta_{1}^{\rm min} = -\eta_{\rm max}$ and $\eta_{s}^{\rm max} = \eta_{\rm max}$, where $\eta_{\rm max}$ is the global maximum value of the space-time rapidity.
One could, for instance, identify $\eta_{\rm max}$ with the beam rapidity.

The subvolume characterizing the physical size of slice $j$ is
\eq{
V_j = \int_{x \in [\eta_{j}^{\rm min},\eta_{j}^{\rm max}]} d \sigma_\mu(x) \, u^{\mu} (x).
}

The key assumption in the following is that each subvolume $V_j$ is sufficiently large for it to be in the thermodynamic limit.
Or in other words, $V_j \gg \xi^3$ for each $i$ where $\xi$ is any relevant correlation length.
If that is the case, one can neglect the surface effects, namely the interactions between particles from different subvolumes.
Mathematically speaking, this implies a scaling $\mathcal{Z}^{\rm HRG}(T,V_j,\bvar \mu) \sim e^{V_j}$~[or, equivalently, $\ln {Z}^{\rm HRG}(T,V_j,\bvar \mu) \sim V_j$] for $V_j \gg \xi^3$.
Also, the total partition function factorizes into a product of partition functions for each of the subvolumes:
\eq{\label{eq:Zeta}
Z_{\rm tot}^{\rm gce, unif} \sim \prod_{j = 1}^{s} \mathcal{Z}^{\rm HRG} (T,V_j,\bs \mu), \quad V_j \gg \xi^3,\\
\label{eq:logZeta}
\ln Z_{\rm tot}^{\rm gce, unif} \simeq \sum_{j = 1}^{s} \ln \mathcal{Z}^{\rm HRG} (T,V_j,\bs \mu), \quad V_j \gg \xi^3.
}

The form of Eq.~\eqref{eq:logZeta} allows us to relax the assumption of the constancy of thermal parameters.
Let us now assume that the intensive thermal parameters depend on the space-time rapidity $\eta_s$.
This implies that each of the rapidity slices is characterized by its own set of values of the thermal parameters, i.e. in Eqs.~\eqref{eq:Zeta}, \eqref{eq:logZeta} one has
$T \to T_i$ and $\bvar \mu \to \bvar \mu_i$:
\eq{\label{eq:Zetadep}
Z_{\rm tot}^{\rm gce} \sim \prod_{j = 1}^{s} \mathcal{Z}^{\rm HRG} (T_j,V_j,\bs \mu_j), \quad V_j \gg \xi^3,\\
\label{eq:logZetadep}
\ln Z_{\rm tot}^{\rm gce} \simeq \sum_{j = 1}^{s} \ln \mathcal{Z}^{\rm HRG} (T_j,V_j,\bs \mu_j), \quad V_j \gg \xi^3.
}

Let us denote the hadron multiplicities in a subvolume $j$
by $\hat{N}_{j} = \{N_{j,i}\}_{i=1}^f$.
The multiplicity distribution $\hat{N}_{j}$ is given by the corresponding multiplicity distribution of the HRG model with thermal parameters of the given subvolume, i.e. $P^{\rm gce}\left(\hat{N}_{j} \right) = P^{\rm hrg}(\hat{N}_{j}; T_j,V_j,\bs \mu_j)$.
Due to the fact that we neglected all correlations between particles from the different subvolumes, the multiplicity distribution of $\hat{N}_{j}$ is independent of the multiplicity distributions in all other subvolumes.
The probability distribution for multiplicities $\left\{ \hat{N}_{j} \right\}_{j=1}^s$ across all subvolumes thus factorizes as follows:
\eq{\label{eq:Pgceall}
P^{\rm gce}\left(\left\{ \hat{N}_{j} \right\}_{j=1}^s \right) = \prod_{j = 1}^{s} P^{\rm hrg}(\hat{N}_{j}; T_j,V_j,\bs \mu_j).
}
The factorization in Eq.~\eqref{eq:Pgceall} will no longer hold once we introduce exact global conservation of conserved charges.

The momentum distribution of hadron species $i$ emitted from a rapidity slice $j$ reads
\eq{\label{eq:CFrap}
E_p \, \frac{dN_{j,i}}{d^3p} = \int_{x \in [\eta_{j}^{\rm min},\eta_{j}^{\rm max}]} d \sigma_\mu (x) \, p^\mu \, f_{j,i}(x, p)~
}
with
\eq{
f_{j,i}(x, p) = \frac{\lambda_i^{\rm int}(T_j,\bs \mu_j)}{(2\pi)^3} \, \exp\left(\frac{\mu_{j,i} - p^\mu  u_\mu(x)}{T} \right).
}
Here $\mu_{j,i} = b_i \mu_{B,j} + q_i \mu_{Q,j} + s_i \mu_{S,j}$.

\subsection{Exact global conservation laws}

Let us now incorporate the effect of exact global conservation of conserved charges.
As we work in the thermodynamic limit, $V_j \gg \xi^3$, the exact conservation will not affect the mean multiplicities due to the thermodynamic equivalence of statistical ensembles.
However, as the thermodynamic equivalence does not extend to  fluctuations, the fluctuation observables will be affected by the exact conservation, no matter how large the system is.

The total values of the globally conserved baryon number, electric charge, and strangeness coincide with the GCE mean values due to the thermodynamic equivalence of ensembles:
\eq{
\bvar Q_{\rm tot} = \sum_{k=1}^s \mean{\bvar Q_k}^{\rm gce}~.
}

To enforce the global conservation laws on the level of multiplicity distributions one has to project out all microstates that violate the global conservations laws from the grand-canonical partition function.
This is achieved by introducing a Kronecker delta  into the grand partition function~\eqref{eq:Zetadep} of the entire system:
\eq{\label{eq:Zce}
Z_{\rm tot}^{\rm ce} & ~ \sim ~ \prod_{j = 1}^{s} \, \sum_{\bvar Q_j} \, e^{\bs \mu_j \cdot \bvar Q_j} \, Z^{\rm HRG} (T_j,V_j,\bvar Q_j)\,
\nonumber \\
&  \qquad \times \delta \left(\bvar Q_{\rm tot} - \sum_{k=1}^s \bvar Q_k \right)~.
}

The presence of the delta function in Eq.~\eqref{eq:Zce} breaks the factorization of multiplicity distributions in different rapidity slices. The joint multiplicity distribution reads
\eq{
\label{eq:PNjoint}
P^{\rm ce}\left(\left\{ \hat{N}_{j} \right\}_{j=1}^s \right) & = \prod_{j = 1}^{s} \,  P^{\rm hrg}(\hat{N}_{j}; T_j,V_j,\bs \mu_j) \nonumber \\
&  \qquad \times \delta \left(\bvar Q_{\rm tot} - \sum_{i=k}^s \bvar Q_k \right), \\
\label{eq:Qhrg}
\bvar Q_k & = \sum_{i=1}^f \, \bvar q_i \, N_{k,i}~.
}
Here $\bvar q_i = (b_i,q_i,s_i)$ is a vector of conserved charge values carried by hadron species $j$.

\subsection{Sampling the multiplicity distribution}

Here we present a general method for sampling the joint multiplicity distribution~Eq.~\eqref{eq:PNjoint} of hadron numbers in all the subsystems.
The method is based on rejection sampling and it assumes that it is known how to sample the multiplicity distribution of the grand-canonical variant of the HRG model used.
To generate a configuration from the distribution~\eqref{eq:PNjoint}
\begin{enumerate}
    \item Sample $\hat{N}_{j}$ for $j = 1\ldots s$ independently for each subsystem from the grand-canonical variant of an interacting HRG model characterizing each subsystem.
    \item Compute $\sum_{k=1}^s \bvar Q_k$ via Eq.~\eqref{eq:Qhrg}. Accept the configuration if $\bvar Q_{\rm tot} = \sum_{k=1}^s \bvar Q_k$, or go back to step 1 otherwise.
\end{enumerate}

The method is general in the sense that it does not assume anything about the specific HRG model used. It will work both for an ideal and interacting HRG.
It should be noted, however, that the algorithm may become inefficient if the acceptance rate in step~2 becomes low.
This can happen for large systems and multiple conserved charges.
More efficient algorithms can be devised for specific versions of the HRG model, see e.g. a multi-step method of Becattini and Ferroni in Ref.~\cite{Becattini:2004rq}.
We do employ this method in our Monte Carlo simulations in Sec.~\ref{sec:LHC}.

\subsection{Thermal smearing}

The algorithm in the previous section allows to sample hadron multiplicity distributions differentially in space-time rapidity.
The experiments, however, perform measurements in momentum rather than coordinate space, therefore, a transition to momentum space is necessary.
In some cases, such as the Bjorken flow scenario at the highest collision energies, it is possible to identify the space-time rapidity $\eta_s$ with the momentum rapidity $Y$, allowing to study rapidity-dependent hadron distributions without the transition to the momentum space.
Even in this case, however, a degree of smearing between $\eta_s$ and $Y$ is present due to thermal motion.
The boost invariance breaks down at lower collision energies and the problem of space-momentum correlations becomes even more severe.
For these reasons it is necessary to assign each of the hadrons a 3-momentum.
Furthermore, if a subsequent afterburner stage is to be included into the modeling, one has to generate both the spatial and momentum coordinates for each hadron.

The procedure to generate the momenta of all the hadrons is fairly straightforward.
Once the multiplicity distributions $\left\{ \hat{N}_{j} \right\}_{j=1}^s$ for all the rapidity slices have been sampled, the coordinates and momenta of all the hadrons can be generated through the standard Cooper-Frye momentum sampling, applied independently to each hadron in each of the rapidity slices.
Several implementations for this task are available, see e.g.~\cite{Kisiel:2005hn,Chojnacki:2011hb,Shen:2014vra}.
The sampled hadrons should then be provided as input into a hadronic afterburner like UrQMD~\cite{Bass:1998ca,Bleicher:1999xi} or SMASH~\cite{Weil:2016zrk}, if one is used, or a cascade of resonance decays performed to obtain the final state particles that are measured experimentally.
The comparison with data can then be done in the standard way, by computing the observables in a given acceptance as statistical averages.

\section{Excluded volume model for net baryon fluctuations}
\label{sec:evhrg}

To illustrate the developed formalism we shall apply it to net proton and net baryon fluctuations in heavy-ion collisions at energies reachable at LHC and RHIC.
In this section we describe the motivation and the technical details behind an excluded volume HRG model that we use for the analysis.
A reader interested only in the final heavy-ion results may skip to Sec.~\ref{sec:LHC} where these are presented and discussed.

The typical chemical freeze-out temperatures, $T_{\rm ch} \sim
155-160$~MeV at the LHC~\cite{Andronic:2017pug,Becattini:2012xb,Petran:2013lja} and $T_{\rm ch} \sim 160-165$~MeV at the top RHIC energies~\cite{Adamczyk:2017iwn}, are close to the pseudo-critical
temperature of the QCD crossover transition determined by lattice QCD $T_{\rm pc} \simeq
155-160$~MeV~\cite{Bazavov:2018mes,Borsanyi:2020fev} at $\mu_B = 0$.
Lattice QCD predicts that the high-order net baryon cumulants, namely the kurtosis $\chi_4^B/\chi_2^B$ and the hyperkurtosis $\chi_6^B/\chi_2^B$ ratios deviate significantly from the Skellam distribution baseline of the ideal HRG model, where these ratios are equal to unity.
The hyperkurtosis in particular turns negative around $T_{\rm pc}$ which is thought to be related to the remnants of the chiral criticality~\cite{Friman:2011pf} at vanishing light quark masses.
It would certainly be of great interest to verify this theory prediction of a negative $\chi_6^B$ experimentally, which may serve as an experimental evidence for the chiral crossover transition.
The measurement of higher-order net proton fluctuations is planned in future runs at the LHC~\cite{Citron:2018lsq}.

In our previous work~\cite{Vovchenko:2020tsr} we studied this question analytically, in the framework of the subensemble acceptance method~(SAM).
There, the sensitivity of measurements to the equation of state was predicted to be not overshadowed if the measurements are performed in acceptance spanning 1-2 units of rapidity.
However, the entire argument in~\cite{Vovchenko:2020tsr} has been done in the configuration space, relying on perfect momentum-space correlations due to Bjorken flow.
Here we would like to determine how the results will be distorted by the thermal smearing and resonance decays.

To apply the formalism of Sec.~\ref{sec:sampler} we need to employ an interacting HRG model that matches the lattice QCD equation of state and be able to sample the grand-canonical multiplicity distribution of such a model.
Here we take an HRG model with excluded volume interactions in the baryonic sector  -- the EV-HRG model -- which was formulated in Refs.~\cite{Vovchenko:2016rkn,Vovchenko:2017xad} and shown to describe well the lattice data on the diagonal net-baryon susceptibilities at $\mu_B = 0$ at temperatures up to and even slightly above $T_{\rm pc}$.

\subsection{Single-component EV model}

Before discussing the full model let us first consider a single-component excluded volume model in order to introduce the multiplicity sampling procedure.
The grand partition function at fixed temperature $T$, volume $V$, and chemical potential $\mu$ reads
\eq{\label{eq:Zev}
Z^{\rm ev} (T,V,\mu) = \sum_{N=0}^{\infty} \, \frac{\left[(V-bN) \, \phi(T) \, e^{\mu/T}\right]^N}{N!} \, \theta(V - bN)~.
}
Here
\eq{\label{eq:phi}
\phi(T) = \frac{d \, m^2 \, T}{2 \pi^2} \, K_2(m/T)~
}
is an ideal gas density of particle species with degeneracy $d$ and mass $m$ at vanishing chemical potential. 
$K_2$ is the modified Bessel function of the second kind.

Equation~\eqref{eq:Zev} defines the multiplicity distribution of the EV model, giving the following (unnormalized) probability function
\eq{\label{eq:Pev}
\tilde{P}^{\rm ev}(N;T,V,\mu) = \frac{\left[(V-bN) \, \phi(T) \, e^{\mu/T}\right]^N}{N!} \, \theta(V - bN)~.
}

In the thermodynamic limit, $N \to \infty$, the particle density $n^{\rm ev}(T,\mu) = \mean{N}^{\rm ev} / V$ is determined by the maximum term in Eq.~\eqref{eq:Zev}. 
Maximizing $\tilde{P}^{\rm ev}$ with respect to $N$ gives a transcendental equation defining $n^{\rm ev}(T,\mu)$:
\eq{\label{eq:transc}
\frac{bn^{\rm ev}}{1-bn^{\rm ev}} e^{\frac{bn^{\rm ev}}{1-bn^{\rm ev}}} = b \, \phi(T) \, e^{\mu/T}.
}
The solution to Eq.~\eqref{eq:transc} is given in terms of the Lambert W function~(see Ref.~\cite{Taradiy:2019taz} for details):
\eq{
\frac{bn^{\rm ev}}{1-bn^{\rm ev}} = W\left[ b\,\phi(T)\,e^{\mu/T} \right],
}
or
\eq{\label{eq:nev}
n^{\rm ev}(T,\mu) = \frac{W\left[ b\,\phi(T)\,e^{\mu/T} \right]}{b \left\{1 + W\left[ b\,\phi(T)\,e^{\mu/T} \right] \right\}}~.
}

The pressure reads
\eq{
p^{\rm ev}(T,\mu) = \frac{Tn^{\rm ev}}{1-bn^{\rm ev}} = \frac{T}{b} \, W\left[ b\,\phi(T)\,e^{\mu/T} \right]~.
}

\subsubsection{Dimensionless form}

In the EV model it is possible to replace the three thermal parameters $(T,V,\mu)$ and the excluded volume parameter $b$ by two dimensionless quantities, namely
a reduced volume $\tilde{V} \equiv V / b$ and a parameter $\varkappa \equiv b \, \phi(T) \, e^{\mu/T}$ that characterizes the strength of repulsive interactions.
The probability distribution~\eqref{eq:Pev} then takes the form
\eq{\label{eq:Pdimless}
\tilde{P}^{\rm ev}(N;\tilde{V}, \varkappa) = \frac{\left[(\tilde{V}-N) \, \varkappa \right]^N}{N!} \, \theta(\tilde{V} - N)~.
}
The mean particle number reads
\eq{\label{eq:Ndimless}
\mean{N}^{\rm ev} = \tilde{V} \, \frac{W(\varkappa)}{1 + W(\varkappa)}~.
}

This reduced form implies that the multiplicity distribution is fully specified if the values of parameters $\tilde{V}$ and $\varkappa$ are known.

\subsubsection{Cumulants of particle number distribution}

Cumulants of the particle number distribution in the EV model can be evaluated from the probability distribution function~\eqref{eq:Pdimless}.
The $n$th moment reads
\eq{\label{eq:momentsum}
\mean{N^n} = \frac{\displaystyle \sum_{N=0}^{\lfloor \tilde{V} \rfloor} N^n \, \tilde{P}^{\rm ev}(N)}{\displaystyle \sum_{N=0}^{\lfloor \tilde{V} \rfloor} \tilde{P}^{\rm ev}(N)}~.
}
The sums over $N$ are finite due to the presence of the $\theta$ function in Eq.~\eqref{eq:Pdimless}. 
Thus, for finite $\tilde{V}$, they can be carried out explicitly.
The cumulants can be expressed in terms of the moments as
\eq{\label{eq:cumusum}
\kappa_n[N] = \sum_{k=1}^n \, (-1)^{k-1} \, (k-1)! \, B_{n,k} (\mean{N}, \ldots, \mean{N^{n-k+1}}).
}
Here $B_{n,k}$ are the partial Bell polynomials.

Explicit expressions for $\kappa_n[N]$ can be obtained in the thermodynamic limit, $\tilde{V} \to \infty$.
This is achieved through the cumulant generating function
\eq{
G_N(t) \equiv \ln \mean{e^{tN}}~.
}
The $t$-dependent mean value $\mean{N}^{\rm ev}(t)$ is obtained from Eq.~\eqref{eq:Ndimless} by a substitution $\varkappa \to \varkappa \, e^t$:
\eq{\label{eq:Ndimlesst}
\mean{N(t)}^{\rm ev} = \tilde{V} \, \frac{W(\varkappa \, e^t)}{1 + W(\varkappa e^t)}~.
}
Equation~\eqref{eq:Ndimlesst} corresponds to the first cumulant. The higher-order cumulants are obtained by differentiating $\mean{N(t)}^{\rm ev}$ with respect to $t$.
The results up to fourth order read
\eq{\label{eq:k1TDL}
\kappa_1^{\rm ev}[N] & = \tilde{V} \, \frac{W(\varkappa)}{1 + W(\varkappa)}\,,\\
\label{eq:k2TDL}
\kappa_2^{\rm ev}[N] & = \tilde{V}\,\frac{W(\varkappa)}{[1+W(\varkappa)]^3}\,,\\
\label{eq:k3TDL}
\kappa_3^{\rm ev}[N] & = \tilde{V}\,\frac{W(\varkappa)\,[1-2\,W(\varkappa)]}{[1+W(\varkappa)]^5}\,,\\
\label{eq:k4TDL}
\kappa_4^{\rm ev}[N] & = \tilde{V}\, \frac{W(\varkappa)\,\{1-8\,W(\varkappa) + 6\, [W(\varkappa)]^2\}}{[1+W(\varkappa)]^7}\,.
}
It follows that all cumulant ratios in the EV model depend exclusively on the value of a single parameter $\varkappa$ in the thermodynamic limit.

\subsection{Sampling the excluded volume model}
\label{sec:samplingsingle}

To sample particle numbers from the probability distribution~\eqref{eq:Pev} of the EV model we will use a rejection sampling technique.
First we sample $N$ from an auxiliary envelope distribution $\tilde{P}^{\rm aux}(N;T,V,\mu)$, which we take to be a Poisson distribution centered around $\mean{N}^{\rm ev}$:
\eq{\label{eq:Paux}
\tilde{P}^{\rm aux}(N;T,V,\mu) = \frac{\left(\mean{N}^{\rm ev}\right)^N}{N!} \, \theta(V - bN)~.
}
Here $\mean{N}^{\rm ev} \equiv n^{\rm ev}(T,\mu) \, V$ with $n^{\rm ev}$ defined by Eq.~\eqref{eq:nev}.
The theta function ensures that the packing limit is not violated, i.e. if for a value $N$ sampled from the Poisson distribution one has $V - bN < 0$, this value is rejected.

To correct for the difference between $\tilde{P}^{\rm aux}(N)$ and $\tilde{P}^{\rm ev}(N)$ we apply rejection sampling for each value of $N$ sampled from $\tilde{P}^{\rm aux}$.
First, we define a weight factor $w(N)$ as the ratio between the true and auxiliary multiplicity distributions:
\eq{\label{eq:wdef}
w(N) & \equiv \frac{\tilde{P}^{\rm ev}(N)}{\tilde{P}^{\rm aux}(N)}
\nonumber
\\
& = \left[ \frac{(1-b\,n) \, \phi(T) \, e{^{\mu/T}}}{n^{\rm ev}} \right]^N~.
}
Here $n \equiv N / V$.
The number $N$ sampled from $\tilde{P}^{\rm aux}(N)$ shall be accepted if $\eta < w(N) / w_{\rm max}$ where $w_{\rm max}$ is the maximum possible value of $w(N)$ and $\eta$ is a random number uniformly distributed in an interval $[0,1]$.

To determine $w_{\rm max} \equiv w(N_{\rm max})$ let us rewrite Eq.~\eqref{eq:wdef} as
\eq{
w(N) = \left[ \frac{1-b\,n}{1-b\,n^{\rm ev}}\, e^{\frac{bn^{\rm ev}}{1-bn^{\rm ev}}} \right]^N
}
where we used Eq.~\eqref{eq:transc}.
$N_{\rm max}$ is determined from $\partial \, w(N) / \partial N = 0$. One obtains an equation
\eq{
\ln \left[ \frac{1-b\,n^{\rm max}}{1-b\,n^{\rm ev}}\, e^{\frac{bn^{\rm ev}}{1-bn^{\rm ev}}} \right] = \frac{b n^{\rm max}}{1-bn^{\rm max}}.
}
The solution to the above equation is $n^{\rm max} = n^{\rm ev}$, i.e. the weight is maximized at the mean value of $N$ in the thermodynamic limit\footnote{Note that $\partial \, w(N) / \partial N = 0$ may generally correspond either to a minimum or a maximum of $w(N)$. The particular case can be clarified by analyzing the second derivative of $w(N)$ with respect to $N$. We checked that $\partial \, w(N) / \partial N = 0$ corresponds to the maximum of $w(N)$ if $b > 0$.
Thus, $\tilde{P}^{\rm aux}(N)$ is an envelope of $\tilde{P}^{\rm ev}(N)$.
}.
$w_{\rm max}$ reads
\eq{
w_{\rm max} = \exp\left[{\frac{bn^{\rm ev}}{1-bn^{\rm ev}}\mean{N^{\rm ev}}}\right]~.
}

In numerical calculations it is more convenient to work directly with normalized weights:
\eq{
\tilde{w}(N) & \equiv \frac{w(N)}{w_{\rm max}} \nonumber \\
& = \left[ \frac{1-b\,n}{1-b\,n^{\rm ev}}\right]^N \, \exp\left[{\frac{bn^{\rm ev}}{1-bn^{\rm ev}}(N - \mean{N^{\rm ev}})}\right]~.
}

The reduced weight in terms dimensionless variables $\tilde{V}$ and $\varkappa$ reads
\eq{\label{eq:wdimless}
\tilde{w}(N) = \left[ \frac{\tilde{V}-N}{\tilde{V}-\mean{N}^{\rm ev}}\right]^N \, \exp\left[\mean{N}^{\rm ev}\,\frac{N - \mean{N}^{\rm ev}}{\tilde{V}-\mean{N}^{\rm ev}}\right]~,
}
where $\mean{N}^{\rm ev}$ is given by Eq.~\eqref{eq:Ndimless}.

The sampling procedure described here is similar to the Monte Carlo EV model analysis performed in Ref.~\cite{Vovchenko:2018cnf}, with one distinction.
In Ref.~\cite{Vovchenko:2018cnf} an importance sampling technique was employed, where each generated event was accepted with a weight $\tilde{w}(N)$.
Here, instead, all accepted events have the same weight, but their sampling involves an additional rejection  step with respect to the weights $\tilde{w}(N)$.

\subsubsection*{Testing the sampling procedure}

To test the sampling procedure described above we take $\varkappa = 0.03$ and perform Monte Carlo sampling for different values of $\tilde{V}$.
The choice $\varkappa = 0.03$ is motivated by the fact that this value is obtained in the EV-HRG model with baryonic excluded volume $b = 1$~fm$^3$ at $T = 160$~MeV and $\mu_B = 0$~\cite{Vovchenko:2017xad}, therefore, the exercise approximately corresponds to sampling the baryon multiplicity distribution in the vicinity of the QCD chiral crossover transition where the EV-HRG model approximates well the QCD cumulants of the net baryon distribution.

\begin{figure}[t]
  \centering
  \includegraphics[width=.49\textwidth]{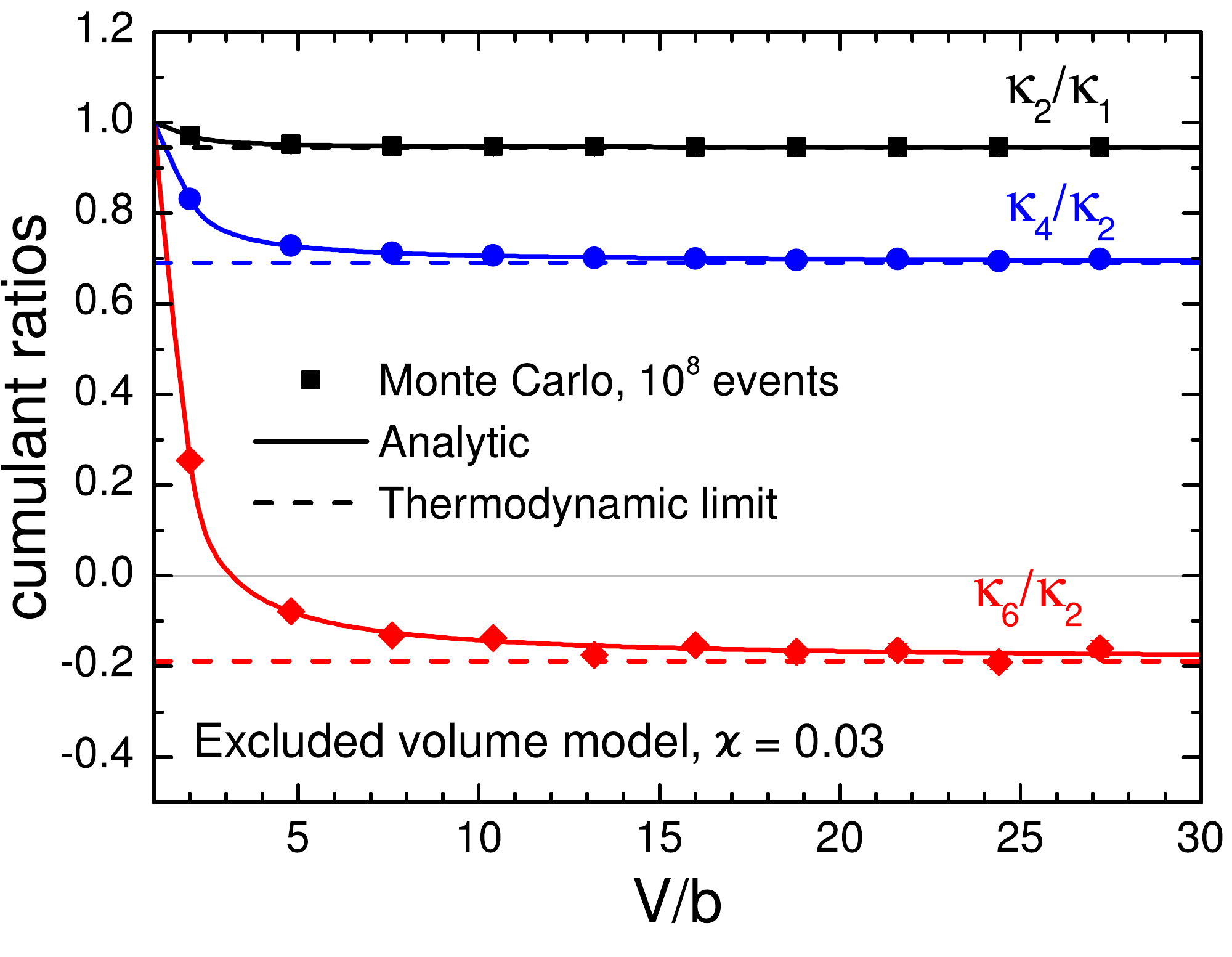}
  \caption{
   The behavior of cumulant ratios $\kappa_2/\kappa_1$~(black), $\kappa_4/\kappa_2$~(blue), and $\kappa_6/\kappa_2$~(red) in a single component grand-canonical excluded volume model as a function of the reduced volume $\tilde{V} \equiv V/b$.
   Calculations are performed through a Monte Carlo sampling of $10^8$ events~(symbols) and analytically via Eqs.~\eqref{eq:momentsum},~\eqref{eq:cumusum}~[solid lines].
   The horizontal dashed lines correspond to cumulant ratios in the thermodynamic limit.
  }
  \label{fig:sampling_test}
\end{figure}

We sample $10^8$ numbers at each value of $\tilde{V}$ and calculate cumulants of the resulting particle number distribution up to $\kappa_6$.
Figure~\ref{fig:sampling_test} depicts the resulting $\tilde{V}$-dependence of the scaled variance $\kappa_2/\kappa_1$, kurtosis $\kappa_4/\kappa_2$, and hyperkurtosis $\kappa_6/\kappa_2$~(symbols).
The solid lines correspond to an analytic calculation of these ratios via a direct summation over all probabilities~[Eqs.~\eqref{eq:momentsum}, \eqref{eq:cumusum}].
The Monte Carlo calculations agree with the analytic expectations at all studied values of $\tilde{V}$, validating the sampling method.

Figure~\ref{fig:sampling_test} allows also to establish when the condition $V \gg \xi^3$ is reached. This is signalled by the approach of the cumulant ratios to their expected values in the thermodynamic limit~[Eqs.~\eqref{eq:k1TDL}-\eqref{eq:k4TDL}], shown in Fig.~\ref{fig:sampling_test} by the horizontal dashed lines.
Cumulant of a higher order generally requires larger values of $\tilde{V}$ to reach the thermodynamic limit, reflecting the fact that higher cumulants are 
more sensitive to
the correlation length $\xi$.
We observe that cumulant ratios up sixth order are within few percent or less of the thermodynamic limit for $\tilde{V} \gtrsim 20$.
The cumulants then scale linearly with the volume for larger values of $\tilde{V}$.
The value $\tilde{V} \simeq 20$ thus establishes a lower bound on the physical volume of a single rapidity slice for the subensemble sampler in Sec.~\ref{sec:sampler} to be applicable.

\subsection{EV-HRG model}

Having established the baryon multiplicity sampling procedure in a single-component case, we now turn to the full model.
Quantitative applications to heavy-ion fluctuation observables require an equation of state with hadron and resonance degrees of freedom matched to first-principle lattice QCD equation of state.
For the purposes of net baryon and net proton fluctuations studied here we employ a variant of an excluded volume hadron resonance gas~(EV-HRG) model introduced in Refs.~\cite{Vovchenko:2016rkn,Satarov:2016peb}.
The repulsive EV interactions are introduced for all baryon-baryon and antibaryon-antibaryon pairs in the EV-HRG model, with a common value $b$ of the EV parameter for all these pairs.

The pressure in the EV-HRG model is partitioned into a sum of meson, baryon and antibaryon contributions
\eq{\label{eq:pevhrg}
p = p_M + p_B+ p_{\bar{B}}.
}
Here
\eq{\label{eq:pMev}
p_{M} & = T \, n_M^{\rm id}(T,\bvar \mu)~, \\
\label{eq:pBev}
p_{B(\bar{B})} & = T \, n_{B(\bar{B})}^{\rm id}(T,\bvar \mu)
\exp\left( \frac{- b \, p_{B(\bar{B})}}{T} \right)~.
}
$n_M^{\rm id}$ and $n_{B(\bar{B})}^{\rm id}$ correspond to cumulative number densities of mesons and (anti)baryons in the ideal HRG limit~($b \to 0$):
\eq{
n_{M,B,\bar{B}}^{\rm id}(T,\bvar \mu) & = \sum_{i \in M,B,\bar{B}} \, n^{\rm id}_i (T, \mu_i)\,\\
n^{\rm id}_i (T, \mu_i) & = \frac{d_i m_i^2 T}{2\pi^2} \, K_2(m_i/T) \, e^{\mu_i/T}.
}
Here $\mu_i = \bvar q_i \cdot \bs \mu$ is the chemical potential of particle species $i$.

The expression~\eqref{eq:pBev} can be rewritten in terms of the Lambert W function in close to analogy to Eq.~\eqref{eq:Pev} of the single-component EV model:
\eq{
p_{B(\bar{B})} = \frac{T}{b} \, W[b \, n_{B(\bar{B})}^{\rm id}(T,\bvar \mu)]~.
}

The particle number densities of individual hadrons species are calculated as derivatives of the pressure with respect to the corresponding chemical potential $n^{\rm ev}_i = \partial p / \partial \mu_i$.
The mean multiplicities $\mean{N_i}^{\rm ev} \equiv V \, n^{\rm ev}_i$ in the grand-canonical EV-HRG model read
\eq{\label{eq:evhrgmesons}
\mean{N_i}^{\rm ev} & = V \, n^{\rm id}_i (T, \mu_i), \qquad i \in M~,\\
\label{eq:evhrgbaryons}
\mean{N_i}^{\rm ev} & = \frac{V}{b} \, \frac{W[\varkappa_{B(\bar{B})}]}{1 + W[\varkappa_{B(\bar{B})}]} \, \frac{n^{\rm id}_i (T, \mu_i)}{n_{B(\bar{B})}^{\rm id}(T,\bvar \mu)}, \nonumber \\
& = V \, n^{\rm id}_i (T, \mu_i) \, \frac{W[\varkappa_{B(\bar{B})}]}{\varkappa_{B(\bar{B})} \left\{1 + W[\varkappa_{B(\bar{B})}]\right\}},
\quad i \in B(\bar{B})~.
}
Here $\varkappa_{B(\bar{B})} \equiv b \, n_{B(\bar{B})}^{\rm id}(T,\bvar \mu)$.
The mean multiplicities of mesons coincide with the ideal HRG model baseline.
The multiplicities of (anti)baryons, on the other hand, are suppressed relative to ideal HRG due to EV interactions. This is quantified by a factor in the r.h.s of Eq.~\eqref{eq:evhrgbaryons}.
For $\varkappa_B \simeq 0.03$, a value corresponding to $T = 160$~MeV and $\mu = 0$~(see below), the yields of baryons are suppressed by about 5\%.

\begin{figure*}[t]
  \centering
  \includegraphics[width=.49\textwidth]{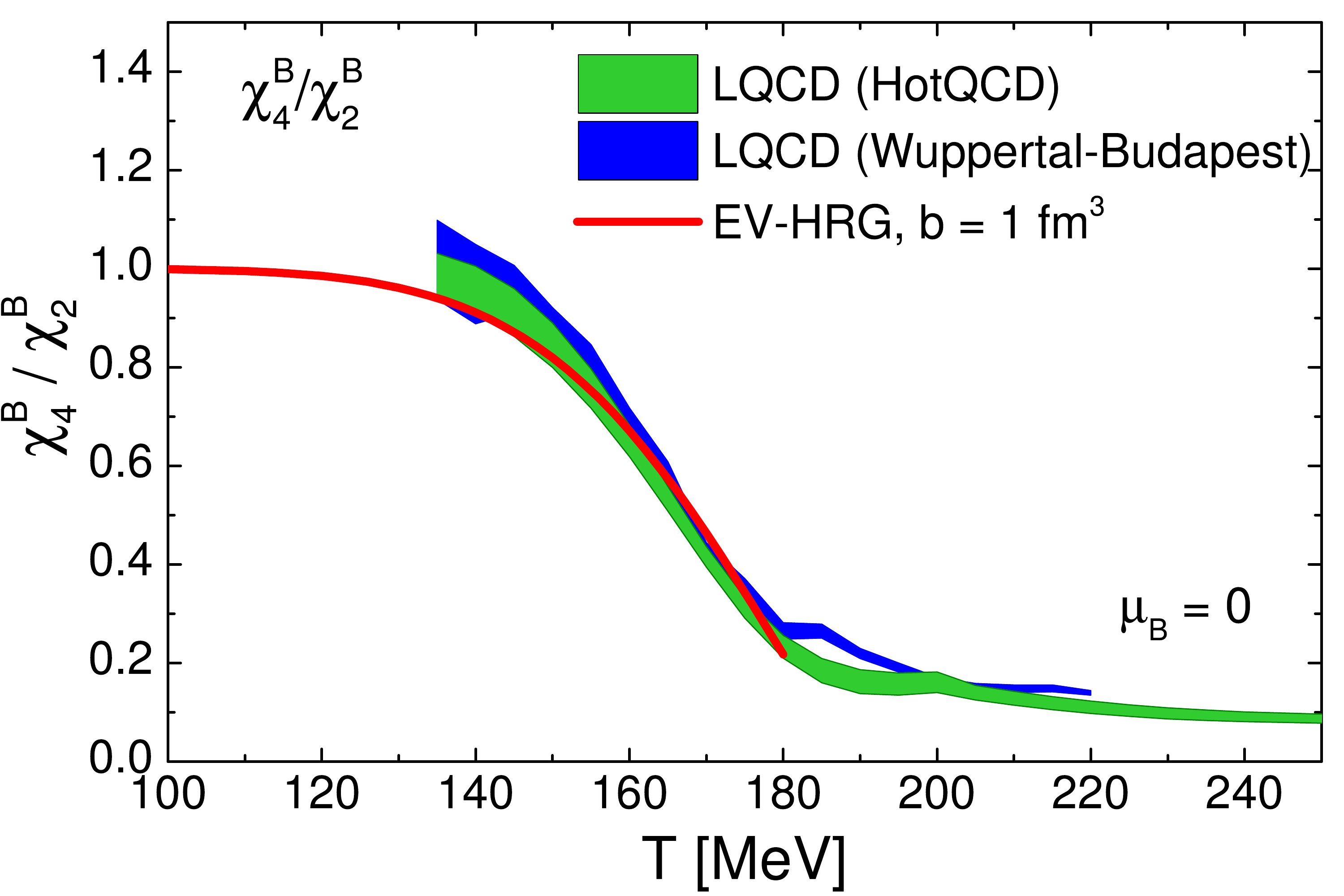}
  \includegraphics[width=.49\textwidth]{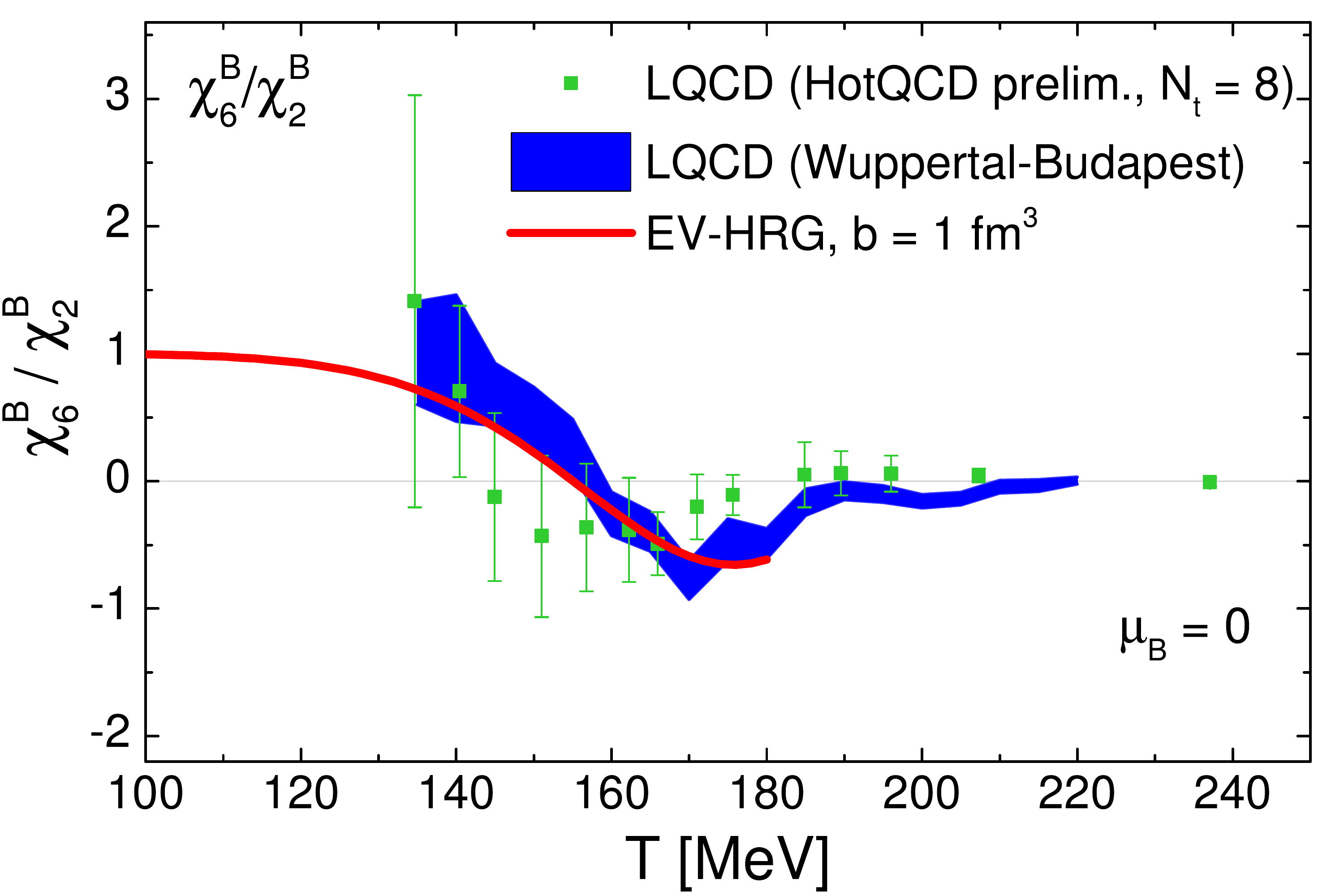}
  \caption{
  Temperature dependence of net baryon susceptibility ratios $\chi_4^B/\chi_2^B$~(left) and $\chi_6^B/\chi_2^B$~(right) evaluated at $\mu_B = 0$ in the EV-HRG model.
  The blue and green bands/symbols depicts lattice QCD results of Wuppertal-Budapest~\cite{Borsanyi:2018grb} and HotQCD~\cite{Bazavov:2017dus} collaborations, respectively.
  }
  \label{fig:chi2evhrg}
\end{figure*}

Equations~\eqref{eq:evhrgmesons} and \eqref{eq:evhrgbaryons} define the factor $\lambda_i^{\rm int}(T,\bs \mu)$ entering the single-particle distribution functions $f_i(x,p)$ for particle species $i$ in the Cooper-Frye formula, Eqs.~\eqref{eq:CF} and~\eqref{eq:CFrap}.
For mesons, $i \in M$, one has $\lambda_i (T,\bs \mu) = 1$. 
For (anti)baryons
\eq{
\lambda_i^{\rm int} (T, \mu) = \frac{W[\varkappa_{B(\bar{B})}]}{\varkappa_{B(\bar{B})} \left\{1 + W[\varkappa_{B(\bar{B})}]\right\}},
\quad i \in B(\bar{B})~.
}

The EV-HRG model has been studied in Refs.~\cite{Vovchenko:2017xad,Vovchenko:2017drx} in the context of lattice QCD results on diagonal net baryon susceptibilities and Fourier coefficients of net baryon density at imaginary chemical potentials. 
Reasonable description of these observables at temperatures close to $T_{\rm pc}$ has been obtained for $b = 1$~fm$^3$, corresponding to $\varkappa_B \simeq 0.03$.
We employ this value of $b$ in the present analysis.
Figure~\ref{fig:chi2evhrg} depicts the temperature dependence of kurtosis $\kappa_4 / \kappa_2$ and hyperkurtosis $\kappa_6 / \kappa_2$ of net baryon fluctuations at vanishing temperatures.
The calculations are compared with the lattice data of  Wuppertal-Budapest~(blue bands)~\cite{Borsanyi:2018grb} and HotQCD~(green bands and symbols)~\cite{Bazavov:2017dus} collaborations.
The model is in quantitative agreement with the lattice data for these two quantities up to $T \simeq 180$~MeV.
This implies that net-baryon distribution of the EV-HRG model in this temperature range closely resembles that of QCD, at least on the level of sixth leading cumulants.
And while this does not necessarily imply that EV interactions is the correct physical mechanism behind the behavior of net baryon susceptibilities, we view the EV-HRG model to be an appropriate tool for the purpose of analysis net baryon and net proton cumulants in heavy-ion collisions.

The sampling procedure in Sec.~\ref{sec:samplingsingle} can be generalized for the EV-HRG model that has multiple hadron components.
We note that the system in the EV-HRG model is partitioned into three independent subsystems, mesons, baryons, and antibaryons, see Eq.~\eqref{eq:pevhrg}.
Therefore, the sampling of the grand-canonical multiplicities proceeds independently for each of the three subsystems.
The multiplicities of the non-interacting mesons are sampled from the Poisson distribution, in the same manner as in the ideal HRG.
The joint probability distribution of numbers of all the baryon species, on the other hand, reads
\eq{
\tilde{P}^{\rm ev}_{B(\bar{B})}(\{N_i\};T,\bvar \mu) & = \left[\prod_{i \in B(\bar{B})} \frac{\left[(\tilde{V}-N_{B(\bar{B})}) \, \varkappa_i \right]^{N_i}}{N_i!} \right]  \nonumber \\ 
& \qquad \times \theta(\tilde{V} - N_{B(\bar{B})})~,
}
where, as before, $\tilde{V} = V / b$ and
\eq{
N_{B(\bar{B})} & = \sum_{i \in B(\bar{B})} \, N_i, \\
\varkappa_i & = b \,  n^{\rm id}_i (T, \mu_i)~.
}

The auxiliary envelope distribution for the sampling is a cut multi-Poisson distribution:
\eq{\label{eq:Pauxmulti}
\tilde{P}^{\rm aux}_{B(\bar{B})}(\{N_i\};T,\bvar \mu) = \left[\prod_{i \in B(\bar{B})} \frac{\left(\mean{N_i}^{\rm ev}\right)^{N_i}}{N_i!} \right] \theta\left(\tilde{V} - N_{B(\bar{B})}\right)~.
}
The theta function is introduced to avoid exceeding the packing limit.

Finally, the normalized weight for the rejection sampling step reads
\eq{\label{eq:wmulti}
\tilde{w}_{B(\bar{B})}(\{N_i\}) & = \prod_{i \in B(\bar{B})} \left[ \frac{1-b\,n_{B(\bar{B})}}{1-b\,n^{\rm ev}_{B(\bar{B})}}\right]^{N_i} \nonumber \\
& \qquad \times \exp\left[{\frac{b \, n^{\rm ev}_{B(\bar{B})}}{1-b \, n^{\rm ev}_{B(\bar{B})}}(N_i - \mean{N^{\rm ev}_i})}\right] \nonumber \\
& = \left[ \frac{1-b\,n_{B(\bar{B})}}{1-b\,n^{\rm ev}_{B(\bar{B})}}\right]^{N_{B(\bar{B})}} \nonumber \\
& \qquad \times \exp\left[{\frac{b \, n^{\rm ev}_{B(\bar{B})}}{1-b \, n^{\rm ev}_{B(\bar{B})}}(N_{B(\bar{B})} - \mean{N^{\rm ev}_{B(\bar{B})}})}\right]
}
Here $n_i \equiv N_i / V$ and $n^{\rm ev}_i \equiv \mean{N_i}^{\rm ev} / V$.

The algorithm for sampling the multiplicity distribution of the EV-HRG model is the following:
\begin{enumerate}
    \item Sample the multiplicities $\{N_i\}$ of all baryons from the cut multi-Poisson distribution~\eqref{eq:Pauxmulti}.
    \item Generate a random number $\eta$ from the uniform distribution on the unit interval (0, 1). If $\eta < \tilde{w}_{B}(\{N_i\})$, go to the next step. Otherwise, go back to step 1.
    \item Repeat steps 1-2 in the same fashion to sample the multiplicities of antibaryons.
    \item Sample multiplicities of mesons from the multi-Poisson distribution of the ideal HRG model.
\end{enumerate}

The procedure for generating the multiplicity distribution in the EV-HRG model in various rapidity slices that are constrained by global conservation of conserved charges, as described in Secs.~\ref{sec:sampler} and~\ref{sec:evhrg}, is implemented in an extended version of the \texttt{Thermal-FIST} package~\cite{Vovchenko:2019pjl}.
We use this package in all our calculations.

\section{Fluctuations in heavy-ion collisions at LHC energies}
\label{sec:LHC}

\subsection{The setup}

We apply our formalism to study the rapidity acceptance dependence of fluctuation observables in heavy-ion collisions.
To proceed we need to specify the partition of the space-time rapidity $\eta_s$ axis into fireballs as well as the $\eta_s$ dependence of thermal parameters and volume.

Let us consider Pb-Pb collisions at $\sqrt{s_{\rm NN}} = 2.76$~TeV.
At midrapidity the chemical freeze-out is characterized by vanishing chemical potentials, temperature values $T \simeq 155-160$~MeV and freeze-out volume per rapidity unit $dV/dy \sim 4000-5000$~fm$^3$~\cite{Andronic:2017pug,Vovchenko:2018fmh}.
The simplest possibility then is to assume boost invariance across the entire space-time rapidity range.
In this scenario, the mean total number of particles of given kind in full space, say charged multiplicity $N_{\rm ch}^{\rm 4\pi}$ or number of (anti)baryons $N_{B(\bar{B})}^{\rm 4\pi}$, is then simply given by multiplying the rapidity density at $y = 0$ by the total (space-time) rapidity coverage $-\eta_s^{\rm max} < \eta_s < \eta_s^{\rm max}$, for example
\eq{\label{eq:4piLHCBI}
N_{\rm ch}^{\rm 4\pi} & = 2 \, \eta_s^{\rm max} \, dN_{\rm ch} / dy|_{y=0}
}
for the charged multiplicity.

The question is how to determine $\eta_s^{\rm max}$.
One possibility is to equate this quantity to the beam rapidity $y_{\rm beam} \approx \ln[\sqrt{s_{\rm NN}}/(2m_N)] \simeq 8$.
However, such an estimate is too crude and will overestimate the actual $N_{\rm ch}^{\rm 4\pi}$.
The rapidity density of charged multiplicity measured at $\sqrt{s_{\rm NN}} = 2.76$~TeV by the ALICE collaboration~\cite{Abbas:2013bpa} is consistent with a Bjorken plateau only in a rapidity range $|y| \lesssim 2$, whereas at higher rapidities $dN_{\rm ch} / dy$ drops.
The entire measured rapidity dependence of $dN_{\rm ch} / dy$ is described well by Gaussian with a width $\sigma = 3.86 \pm 0.05$~\cite{Abbas:2013bpa}.
We can use this fact to relate $N_{\rm ch}^{\rm 4\pi}$ and $dN_{\rm ch} / dy|_{y=0}$ in an empirical way:
\eq{\label{eq:4piLHC}
N_{\rm ch}^{\rm 4\pi} & = \int \, dy  \, \exp\left(-\frac{y^2}{2\sigma^2} \right) \, dN_{\rm ch} / dy|_{y=0} \nonumber \\
& \simeq  (9.68 \pm 0.13) \, dN_{\rm ch} / dy|_{y=0}~.
}
Here the error is due to the uncertainty in the value of $\sigma$.
Comparing Eq.~\eqref{eq:4piLHC} with \eqref{eq:4piLHCBI} one obtains
\eq{
\eta_s^{\rm max} = (4.84 \pm 0.07)  \quad \text{for} \quad \sqrt{s_{\rm NN}} = 2.76~\text{TeV}.
}

We shall use a value $\eta_s^{\rm max} = 4.8$ for Pb-Pb collisions at $\sqrt{s_{\rm NN}} = 2.76$~TeV in the following.
We take $T = 160$~MeV and $\bvar \mu = 0$ for all rapidities.
With this choice the model accurately reproduces the rapidity densities of various hadron species at $|y| \lesssim 2$, where the Bjorken plateau is observed in the data~\cite{Abbas:2013bpa}, and provides an accurate estimate of the total hadron multiplicities in full phase space. 
As we have assumed boost invariance across the entire space-time rapidity range, the model does not describe rapidity distributions at $|y| \gtrsim 2$ and thus should not be applied to calculate observables at large rapidities.
However, given the fact that the model does reproduce the $4\pi$ charged multiplicity, it is suitable to describe the influence of global conservation laws on observables computed around midrapidity, $|y| \lesssim 2$.
In the following we focus on these regions around midrapidity.
In a more general study the assumption of boost invariance can be relaxed to incorporate a more accurate description of the forward-backward rapidity regions.

Our model yields a vanishing total net baryon number in the full space.
Essentially, this means that we neglect baryons from the fragmentation regions.
This is similar to a recent study~\cite{Braun-Munzinger:2020jbk} performed in the framework of the ideal HRG model.
There it was estimated that the effect of fragmentation baryons at the LHC does not exceed 6\% for the sixth order net proton cumulant.
We therefore expect the possible influence of the fragmentation region baryons on our results to be small.

We partition the space-time rapidity axis uniformly into slices of width $\Delta \eta_s = 0.1$.
With $\eta_s^{\rm max} = 4.8$ this implies a total of 96 slices. 
The volume of a single slice in 5\% most central collisions is $V_i = dV/dy \, \Delta \eta_s \simeq 400$~fm$^3$.
This value is sufficiently large to ensure that the thermodynamic limit is reached in each of the subvolumes and thus the requirements for the validity of the sampling procedure described in Sec.~\ref{sec:sampler} satisfied.
This also implies that all intensive quantities, such as cumulant ratios, are independent of the value of $V_i$ in this regime, i.e. $V_i$ can be scaled up and down as long as $V_i \gg \xi^3$.
This feature is very useful for the Monte Carlo sampling procedure. 
Indeed, as the statistical error in higher-order cumulants increases with the volume, this error can be minimized by choosing the volume as small as possible.
According to Fig.~\ref{fig:sampling_test}, a value $V_i = 20$~fm$^3$ is sufficiently large to capture all the relevant physics  for cumulants up to sixth order.
For this reason we take $V_i = 20$~fm$^3$ in our Monte Carlo simulations and then linearly scale up the resulting cumulants to match the volume $V_i = 400$~fm$^3$ in 0-5\% Pb-Pb collisions.

We take the EV-HRG model with $b = 1$~fm$^3$. 
As discussed in Sec.~\ref{sec:samplingsingle}, this model provides a reasonable description of high-order net baryon susceptibilities from lattice QCD.
The grand-canonical distribution of hadron multiplicities can be efficiently sampled following the rejection sampling based algorithm described in Sec.~\ref{sec:samplingsingle}.
We take $T_i = 160$~MeV and vanishing chemical potentials, $\bvar \mu_i = 0$, uniformly for all subvolumes along the rapidity axis.

For the net baryon cumulants we shall take into account only the exact conservation of baryon number, which is exactly vanishing, $B = 0$, in all events.
In principle, one should also take into account the exact conservation of electric charge and strangeness.
However, as discussed in Refs.~\cite{Vovchenko:2020tsr,Vovchenko:2020gne}, the influence of these conserved charges on net baryon cumulants is negligible at LHC energies.
Their influence on net proton cumulants is more sizable~\cite{Vovchenko:2020gne} but still expected to be subleading compared to baryon number conservation.
Neglecting the exact conservation of electric charge and strangeness allows to significantly speed up the Monte Carlo event generation, as this strongly reduces the rejection rate associated with exact conservation of multiple conserved charges and allows to gather enough statistics within a reasonable time period to accurately evaluate cumulants up to sixth order.
We do analyze the influence of electric charge and strangeness conservations on 2nd order cumulants of various net-particle distributions in Sec.~\ref{sec:piKL}

Once the joint hadron multiplicity distribution from all the subvolumes has been sampled, we generate the hadron momenta, independently for each hadron. 
To that end we employ the blast-wave model~\cite{Schnedermann:1993ws}, which provides a reasonable description of bulk particle's $p_T$ spectra at LHC~\cite{Abelev:2013vea}. 
The model corresponds to a particlization of a cylindrically shaped fireball~($\sqrt{r_x^2 + r_y^2} \leq r_{\rm max}$), at a constant value of the longitudinal proper time $\tau = \tau_0$. The longitudinal collective motion obeys the Bjorken scaling while the radial velocity scales with the transverse radius, $\beta_r \propto r_\perp^n$.
This corresponds to a flow profile $u^\mu(x) = (\cosh \rho \, \cosh \eta_s, \sinh \rho \, \Vec{e}_{\perp}, \cosh \rho \, \sinh \eta_s)$, where $\rho = \tanh^{-1} \beta_r$, and $\beta_r = \beta_s \zeta^n$ is the transverse flow velocity profile.  Here $\zeta \equiv r_\perp / r_{\rm max}$ is a normalized transverse radius.
The momentum distribution of hadron species with mass $m$ emerging from a $j$th space-time rapidity subvolume is given by
\eq{\label{eq:blastwave}
\frac{dN}{p_T \, d p_T \, dy} & \propto m_T \, \int_{\eta_{j}^{\rm min}}^{\eta_{j}^{\rm max}} 
d \eta \cosh (y - \eta) \, \int_0^1 \zeta \, d \zeta  \nonumber \\
& \quad \times 
e^{-\frac{m_T \, \cosh \rho \, \cosh (y - \eta)}{T}}
\, I_0 \left(\frac{p_T \sinh \rho}{T}\right),
}
Here $m_T = \sqrt{p_T^2 + m^2}$ is the transverse mass, $y = \displaystyle \frac{1}{2} \log \frac{\omega_p - p_z}{\omega_p + p_z}$ is the longitudinal rapidity, and $I_0$ is a modified Bessel function.

\begin{figure}[t]
  \centering
  \includegraphics[width=.49\textwidth]{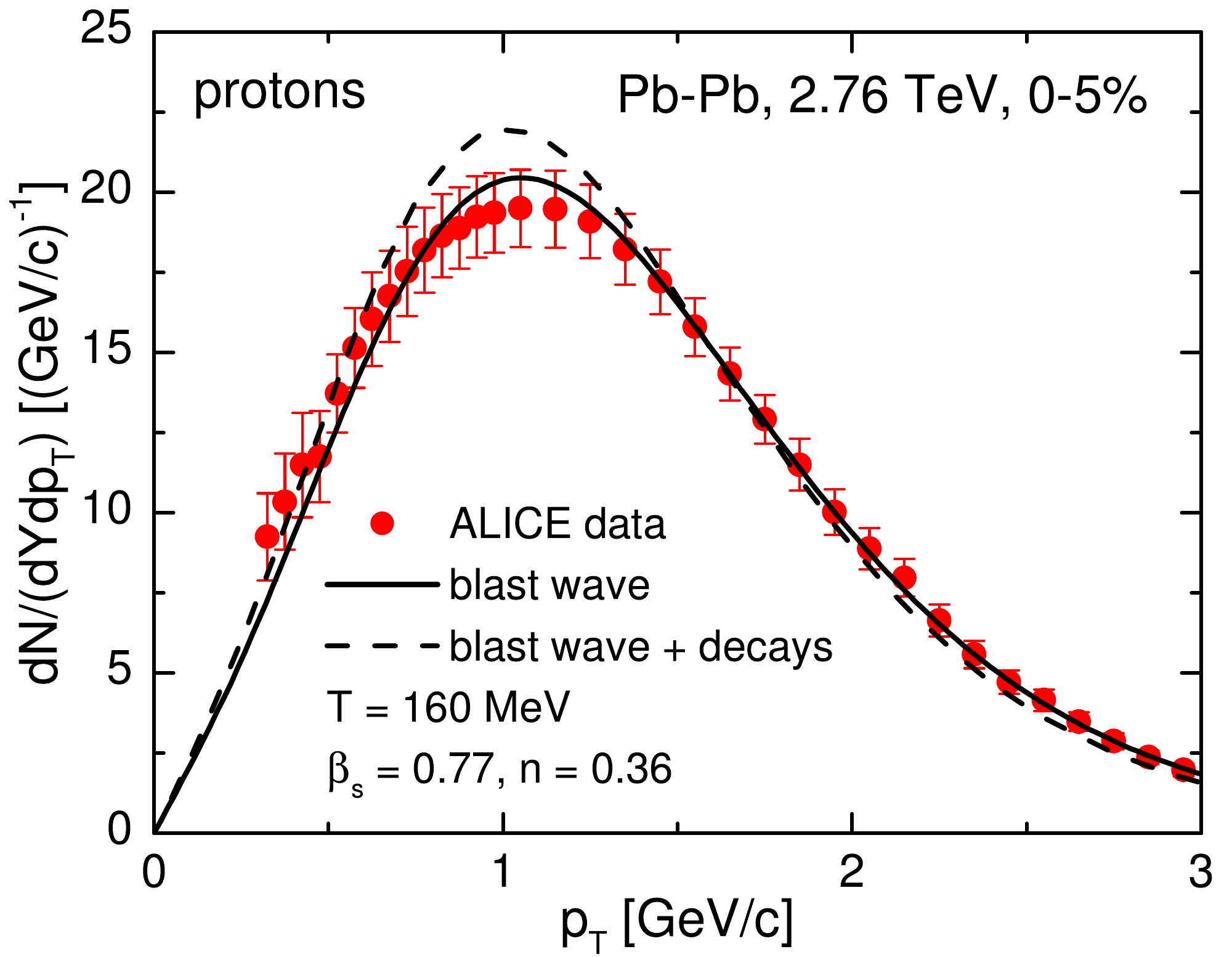}
  \caption{
   The $p_T$ spectrum of protons in 5\% most central Pb-Pb collisions at $\sqrt{s_{\rm NN}} = 2.76$~TeV at midrapidity~($|y| < 0.5$), as measured by the ALICE collaboration~(red symbols)~\cite{Abelev:2013vea} and given by the blast-wave model~($T = 160$~MeV, $\beta_s = 0.77$, $n = 0.36$) with~(dashed line) or without~(solid line) $p_T$ shape modification due to resonance decays.
   The normalization factor of the blast-wave model distribution has been fitted to the data.
  }
  \label{fig:protonsBW}
\end{figure}

The sampling of momenta from the distribution~\eqref{eq:blastwave} is readily implemented in the \texttt{Thermal-FIST} package that we employ.
We are only left with specifying the values of the blast-wave model parameters $\beta_s$ and $n$.
For this purpose we make use of the result of a recent study~\cite{Mazeliauskas:2019ifr}, where the blast-wave model was fitted to experimental data of the ALICE collaboration with account for modification of $p_T$ spectra due to resonance decays.
For 5\% most central Pb-Pb collisions at $\sqrt{s_{\rm NN}} = 2.76$~TeV one has $\beta_s = 0.77$ and $n = 0.36$, which gives a reasonable description of bulk hadron $p_T$ spectra\footnote{One notable exception here are low-$p_T$ pions that are significantly underestimated by the blast-wave model. These pions have no influence on the net baryon fluctuations that we study here.}.
One should note that Ref.~\cite{Mazeliauskas:2019ifr} has extracted a temperature value of $T = 149$~MeV from the $p_T$ spectra fits rather rather than the $T = 160$~MeV value that we use here for fluctuations.
However, the $T = 160$~MeV value shows a similarly good agreement of the blast-wave model proton $p_T$ spectrum  with the data, as the one shown in~\cite{Mazeliauskas:2019ifr} for $T = 149$~MeV.
Figure~\ref{fig:protonsBW} compares the shape of the $p_T$ spectrum of protons as observed in the data~(red symbols)~\cite{Abelev:2013vea} and predicted by the blast-wave model~[Eq.~\eqref{eq:blastwave}] with $T = 160$~MeV, $\beta_s = 0.77$, and $n = 0.36$.
The dashed line in Fig.~\ref{fig:protonsBW} corresponds to blast-wave model spectrum which includes the modification of the proton $p_T$ spectrum due to resonance decays. This effect, computed here via Monte Carlo simulations of decays, only slightly modifies the momentum distribution.

In the final step of the Monte Carlo event generation procedure we perform all strong and electromagnetic decays until only stable hadrons are left.
We generate $10^{10}$ events in total\footnote{Such a large number of events is needed to compute cumulants of sixth order with a sufficiently small statistical uncertainty.} and study the rapidity dependence of various fluctuation observables.
As our analysis only concerns the baryons, to speed-up the Monte Carlo procedure we omit all the primordial mesonic species~(step 4 in the algorithm of Sec.~\ref{sec:samplingsingle}), as these do not affect the behavior of (anti)baryons in any way within the EV-HRG model that we use.

\subsection{Rapidity acceptance dependence of net baryon cumulants}

We start with the rapidity acceptance dependence of net baryon number cumulants.
First, we look at the second cumulant of net baryon fluctuations normalized by the Skellam distribution baseline, $\kappa_2[B-\bar{B}] / \mean{B + \bar{B}}$.
This type of ratio has been extensively studied at LHC energies by the ALICE collaboration~\cite{Acharya:2019izy} for net protons.
This ratio equals unity for the case of a grand-canonical ideal HRG model at any temperature and chemical potentials.
The ratio, however, does exhibit small deviations from unity in the EV-HRG model that we use.
For instance, at $T = 160$~MeV and $\bvar \mu = 0$ the grand-canonical value reads
\eq{\label{eq:c2Skgce}
\left( \frac{\kappa_2[B-\bar{B}]}{\mean{B + \bar{B}}} \right)^{\rm ev,gce} \simeq 0.94~.
}

We note that it is currently challenging to directly compute $\kappa_2[B-\bar{B}] / \mean{B + \bar{B}}$ in lattice QCD, as the denominator $\mean{B + \bar{B}}$ is not a conserved quantity.
Given the good agreement of the EV-HRG model with lattice QCD for the higher-order cumulants, however, we expect QCD to have a similar value to the one given by Eq.~\eqref{eq:c2Skgce}.
An interesting question now is to determine if and how the grand-canonical value in Eq.~\eqref{eq:c2Skgce} is reflected in heavy-ion data.

\begin{figure}[t]
  \centering
  \includegraphics[width=.49\textwidth]{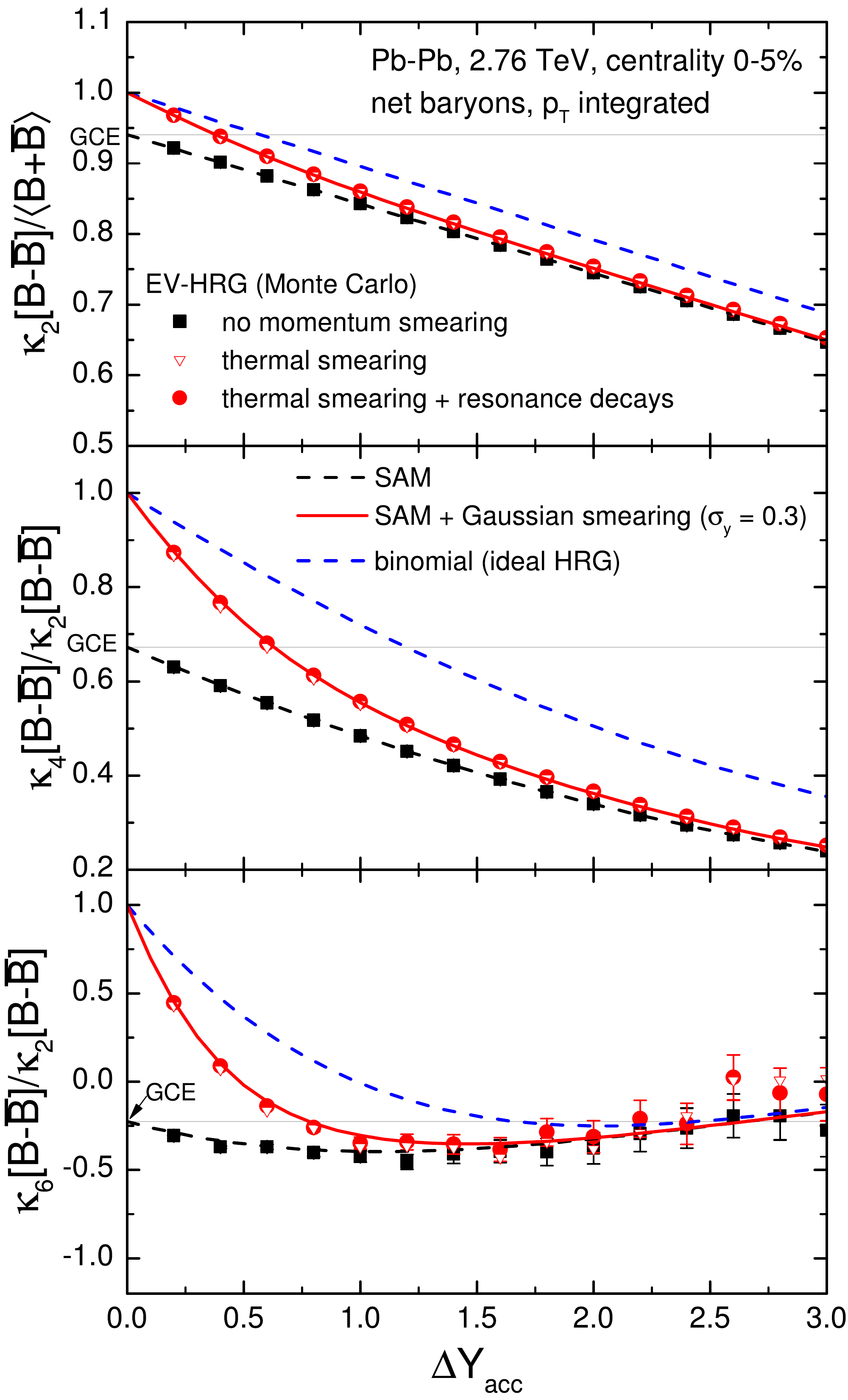}
  \caption{
  Rapidity acceptance dependence of cumulant ratios $\kappa_2/\kappa_2^{\rm Skellam}$ (top), $\kappa_4/\kappa_2$ (middle), and $\kappa_6/\kappa_2$ (bottom) of net baryon distribution in 0-5\% central Pb-Pb collisions at the LHC in an excluded volume HRG model matched to lattice QCD.
  The symbols depict the results of the Monte Carlo event generator, the full black squares correspond to neglecting the momentum smearing, the open red triangles include the thermal smearing at particlization, and the full red circles incorporate the smearing due to both the thermal motion and resonance decays.
  The dashed black lines correspond to the predictions of the SAM framework~\cite{Vovchenko:2020tsr}.
  The solid red lines correspond to adding a Gaussian rapidity smearing on top of the SAM.
  The dashed blue lines correspond to the binomial acceptance, which describes the effects of baryon number conservation in the ideal HRG model limit.
  }
  \label{fig:netbaryonLHC}
\end{figure}

The top panel of Fig.~\ref{fig:netbaryonLHC} depicts the rapidity acceptance 
$\Delta Y_{\rm acc}$
dependence of $\kappa_2[B-\bar{B}] / \mean{B + \bar{B}}$ that results from the Monte Carlo sampling within the subensemble sampler.
Here the acceptance is centered at midrapidity, i.e. particles with rapidity $|y| < \Delta Y_{\rm acc}/2$ are accepted.
The red symbols depict the full result which includes the distortion of hadron momenta due to thermal smearing at particlization and subsequent resonance decays.
The black symbols, on the other hand, correspond to the case when these effects are neglected, i.e. the final kinematical rapidity is taken to be equal to the space-time rapidity coordinate at particlization, $y \equiv \eta_s$.
Comparing the two allows to establish the effect of thermal smearing and resonance decays.
We observe that the Monte Carlo results in the no-smearing case agree with the analytic expectations of the SAM~(black lines).
The SAM baseline for $\kappa_2[B-\bar{B}] / \mean{B + \bar{B}}$ is given by~\cite{Vovchenko:2020tsr,Vovchenko:2020gne}
\eq{\label{eq:SAMc2}
\left( \frac{\kappa_2[B-\bar{B}]}{\mean{B + \bar{B}}} \right)^{\rm SAM} = (1-\alpha) \, \left( \frac{\kappa_2[B-\bar{B}]}{\mean{B + \bar{B}}} \right)^{\rm ev,gce}~.
}
Here $\alpha$ is a fraction of the total volume which corresponds to the acceptance $|\eta_S| < \Delta Y_{\rm acc}/2$:
\eq{
\alpha = \frac{\Delta Y_{\rm acc}}{2\,\eta_s^{\rm max}}~.
}
The agreement of the Monte Carlo points with the SAM is the expected result and serves as a validation of the sampling procedure.

Notable differences between the red~(momentum rapidity) and black~(space-time rapidity) points in Fig.~\ref{fig:netbaryonLHC} appear when the acceptance is sufficiently small, $\Delta Y_{\rm acc} \lesssim 1$.
This is a consequence of the dilution of momentum-space correlations due to thermal motion.
For a very small acceptance, $\Delta Y_{\rm acc} \ll 1$, 
the results converge to the baseline given by the binomial distribution, $\left( \kappa_2[B-\bar{B}] / \mean{B + \bar{B}} \right)^{\rm binom} = 1-\alpha$, shown in Fig.~\ref{fig:netbaryonLHC} by the dashed blue line.
The binomial distribution corresponds to an independent acceptance for all (anti)particles and describes the cumulants of net baryon distribution in the ideal HRG model, where the global baryon conservation constitutes the only source of correlations between baryons~\cite{Bzdak:2012an,Braun-Munzinger:2016yjz,Savchuk:2019xfg}.

The additional momentum smearing due to decays of baryonic resonances is virtually negligible, being completely overshadowed by the thermal smearing. 
This is true not only for the variance, but also for the kurtosis and hyperkurtosis, as seen by comparing the red points~(thermal smearing + resonance decays) with the open red triangles~(thermal smearing only) in all three panels of Fig.~\ref{fig:netbaryonLHC}.
To understand this behavior one can consider e.g. decays $\Delta \to N \pi$.
In such a decay the released momentum is 
split evenly between the two decay products in the resonance center-of-mass frame.
This leads to a larger velocity~(rapidity) smearing of the lighter decay product -- the pion -- whereas the velocity~(rapidity) of nucleon is less affected.
We conclude that the smearing of baryon fluctuations due to resonance decays can be safely neglected.
Note that this statement does not extend to (net-)particle fluctuations involving lighter hadrons such as pions or kaons.
There the effect of resonance decays should be carefully taken into account.

In the Appendix we develop a simplified analytic model to take into account the momentum smearing in net baryon cumulants. 
There we assume that the shift in kinematical rapidity relative to the space-time rapidity is described for all baryon species by a Gaussian smearing.
The red lines in Fig.~\ref{fig:netbaryonLHC} exhibit the results of such a simplified calculation. 
For a Gaussian width of $\sigma_y = 0.3$ the simplified calculations agree very well with the full Monte Carlo results.
Therefore, this model can be used to predict the rapidity dependence of $p_T$-integrated net baryon cumulants without invoking the time-consuming Monte Carlo event generator.

Net-baryon fluctuations in a sufficiently large rapidity acceptance $\Delta Y_{\rm acc} \gtrsim 1$ are accurately described by the analytical SAM baseline~\eqref{eq:SAMc2}.
This conclusion is important, because the SAM makes the connection between the grand-canonical susceptibilities and cumulants constrained by global conservation laws without any additional assumptions regarding the equation of state.
In our previous work~\cite{Vovchenko:2020tsr} where the SAM is introduced, we argued that the SAM is reliable for rapidity acceptances $\Delta Y_{\rm acc} \gtrsim 1$, where the distortion due to thermal smearing is expected to be subleading.
The results obtained in the present work using the EV-HRG model explicitly confirm this.
The grand-canonical $\left( \kappa_2[B-\bar{B}] / \mean{B + \bar{B}} \right)^{\rm ev,gce}$ can therefore be extracted from data by fitting the $\alpha$-dependence of net-baryon fluctuations measured in sufficiently large rapidity acceptance via Eq.~\eqref{eq:SAMc2}.

We turn now to the kurtosis of net baryon fluctuations, $\kappa_4[B-\bar{B}]/\kappa_2[B-\bar{B}]$.
In the grand-canonical ensemble this quantity coincides with the corresponding ratio $\chi_4^B/\chi_2^B$ of the susceptibilities.
The EV-HRG model at LHC energies yields the following value
\eq{\label{eq:chi4chi2}
\frac{\chi_4^B}{\chi_2^B} \simeq 0.67.
}
This is in agreement with lattice QCD continuum estimates of HotQCD ($\chi_4^B/\chi_2^B = 0.65 \pm 0.03$)~\cite{Bazavov:2017dus} and Wuppertal-Budapest ($\chi_4^B/\chi_2^B = 0.69 \pm 0.03$)~\cite{Borsanyi:2018grb} collaborations, taken at the same temperature $T = 160$~MeV.

The rapidity acceptance dependence of $\kappa_4[B-\bar{B}]/\kappa_2[B-\bar{B}]$ is depicted in the middle panel of Fig.~\ref{fig:netbaryonLHC}.
The qualitative behavior of the kurtosis largely mirrors that of the variance.
In the absence of momentum smearing, the Monte Carlo results agree with the analytical SAM baseline of Ref.~\cite{Vovchenko:2020tsr}:
\eq{\label{eq:SAMc4c2}
\left( \frac{\kappa_4[B-\bar{B}]}{\kappa_2[B-\bar{B}]} \right)^{\rm SAM} 
= 
(1-3\alpha \beta) \, \frac{\chi_4^B}{\chi_2^B} - 3 \alpha \beta \, \left(\frac{\chi_3^B}{\chi_2^B}\right)^2~.
}
Here $\beta \equiv 1 - \alpha$.
At LHC energies one has $\chi_3^B/\chi_2^B = 0$, thus, the second term in Eq.~\eqref{eq:SAMc4c2} does not contribute.

With thermal smearing and resonance decays included, the kurtosis deviates from the SAM baseline for $\Delta Y_{\rm acc} \lesssim 1$ and for $\Delta Y_{\rm acc} \ll 1$ tends to the binomial distribution baseline, which at the LHC energies reads $\left( \kappa_4[B-\bar{B}] / \kappa_2[B-\bar{B}] \right)^{\rm binom}_{\rm LHC} = 1-3\alpha \beta$.
For $\Delta Y_{\rm acc} \gtrsim 1$ the full result is described well by the SAM~\eqref{eq:SAMc4c2}.

Finally, we look at the behavior of the hyperkurtosis, $\kappa_6[B-\bar{B}]/\kappa_2[B-\bar{B}]$.
Lattice QCD predicts a sign change of the grand-canonical hyperkurtosis at $\bvar \mu = 0$ in the vicinity of the pseudocritical temperature~(Fig.~\ref{fig:chi2evhrg}).
This qualitative feature is thought to be a signature of the QCD chiral crossover transition~\cite{Skokov:2012ds}. 
Therefore, a corresponding measurement of $\kappa_6[B-\bar{B}]/\kappa_2[B-\bar{B}]$ in heavy-ion collisions at the LHC can potentially serve as the first experimental signature of that transition.
The EV-HRG model reproduces the available lattice QCD data for $\chi_6^B/\chi_2^B$ and gives the following value at $T = 160$~MeV:
\eq{\label{eq:chi6chi2}
\frac{\chi_6^B}{\chi_2^B} \simeq -0.23.
}
This agrees within errors with the continuum estimate of the Wuppertal-Budapest collaboration, $\chi_6^B/\chi_2^B = -0.26 \pm 0.17$~\cite{Borsanyi:2018grb} as well as with $N_\tau = 8$ results of the HotQCD collaboration~\cite{Bazavov:2017dus}  shown in Fig.~\ref{fig:chi2evhrg}.

The lower panel of Fig.~\ref{fig:netbaryonLHC} shows the rapidity acceptance dependence of the hyperkurtosis.
In the absence of momentum smearing, the Monte Carlo results are described by the analytical SAM baseline, which for LHC energies, i.e. for $\bvar \mu = 0$, reads~\cite{Vovchenko:2020tsr}
\eq{\label{eq:SAMc6c2}
\left( \frac{\kappa_6[B-\bar{B}]}{\kappa_2[B-\bar{B}]} \right)^{\rm SAM}_{\rm LHC}
& = 
\left[1-5\alpha \beta (1 - \alpha \beta ) \right] \frac{\chi_6^B}{\chi_2^B} \nonumber \\
& \quad - 10 \alpha (1-2\alpha)^2 \beta \left( \frac{\chi_4^B}{\chi_2^B}\right)^2~.
}

The hyperkurtosis, in the absence of momentum smearing, is sensitive to the grand-canonical value~\eqref{eq:chi6chi2} in acceptances up to $\Delta Y_{\rm acc} \lesssim 1.5$.
For larger acceptances baryon conservation dominates, making it difficult to disentangle between the EV-HRG model and the binomial baseline, given by $\left( \kappa_6[B-\bar{B}] / \kappa_2[B-\bar{B}] \right)^{\rm binom}_{\rm LHC} = 1-15\alpha\beta(1-3\alpha \beta)$.
This was already pointed out in our previous study~\cite{Vovchenko:2020tsr}.
The thermal smearing distorts the signal at small acceptances, $\Delta Y_{\rm acc} \lesssim 0.5$, where the hyperkurtosis is closer to the binomial distribution baseline than it is to the SAM. 
At $0.5 \lesssim \Delta Y_{\rm acc} \lesssim 1.5$, on the other hand, $\kappa_6[B-\bar{B}]/\kappa_2[B-\bar{B}]$ is overshadowed neither by the thermal smearing nor by the baryon number conservation.
We, therefore, argue that a measurement of a hyperkurtosis, which is negative over this entire range may be interpreted as a signal of the chiral crossover\footnote{We note that at $\Delta Y_{\rm acc} \gtrsim 1$ baryon number conservation leads to a negative hyperkurtosis also in the case of the ideal HRG, see the dashed blue line in Fig.~\ref{fig:netbaryonLHC}. Thus it is essential to establish a negative $\kappa_6[B-\bar{B}]/\kappa_2[B-\bar{B}]$ at $\Delta Y_{\rm acc} \lesssim 1$ for the chiral crossover interpretation to be valid.}.

\subsection{Net baryon vs net proton fluctuations}

Our discussion has so far been restricted to cumulants of net baryon distribution.
However, experiments typically cannot measure all baryons, in particular the measurement of neutrons is extremely challenging.
For this reason one usually uses net protons as a proxy for net baryons.
It is natural to expect net protons to carry at least some information about net baryon fluctuations.
In fact, as shown by Kitazawa and Asakawa~\cite{Kitazawa:2011wh,Kitazawa:2012at}, under the assumption of isospin randomization at late stages of heavy-ion collisions, one can reconstruct the cumulants net baryon distribution from the measured factorial moments of  proton and antiproton distributions.

\begin{figure}[t]
  \centering
  \includegraphics[width=.49\textwidth]{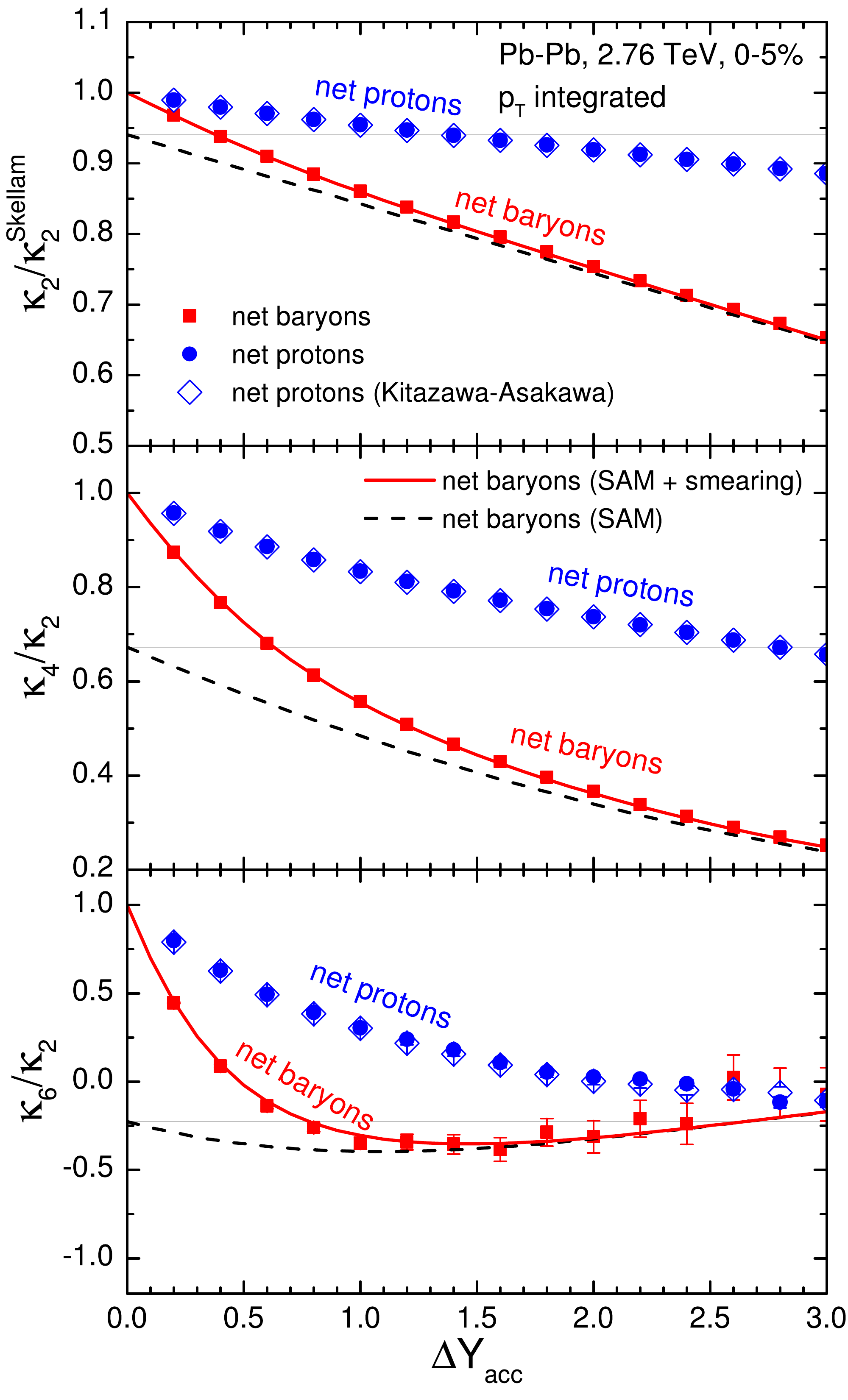}
  \caption{
   Rapidity acceptance dependence of net baryon~(black squares) and net proton~(blue circles) cumulant ratios $\kappa_2/\kappa_2^{\rm Skellam}$ (top), $\kappa_4/\kappa_2$ (middle), and $\kappa_6/\kappa_2$ (bottom) in 0-5\% central Pb-Pb collisions at the LHC in an excluded volume HRG model matched to lattice QCD.
   The open blue diamonds correspond to net proton cumulants evaluated from net baryon cumulants using a binomial-like method of Kitazawa and Asakawa~\cite{Kitazawa:2011wh,Kitazawa:2012at}.
   The black lines correspond to the analytical predictions of the SAM framework with~(solid) and without~(dashed) Gaussian rapidity smearing.
  }
  \label{fig:netbaryonproton}
\end{figure}

However, these considerations do \emph{not} imply that ratios of proton cumulants can be used directly in place of the corresponding ratios of baryon cumulants,
something which has nevertheless been employed in a number of works in the literature~\cite{Albright:2015uua,Fu:2016tey,Bazavov:2020bjn}.
The proton and baryon cumulant ratios do coincide in the free hadron gas limit, where they both trivially reduce to the Skellam baseline, but this does not hold in general case.

Large differences between net proton and net baryon cumulant ratios were reported earlier in Ref.~\cite{Vovchenko:2017ayq} for the van der Waals HRG model in the grand-canonical ensemble.
Here we study these differences in the framework of the EV-HRG model constrained to lattice data and include effects of global baryon conservation and momentum smearing.

Figure~\ref{fig:netbaryonproton} depicts the rapidity acceptance dependence of net baryon~(black squares) and net proton~(blue symbols) cumulant ratios $\kappa_2/\kappa_2^{\rm Skellam}$, $\kappa_4/\kappa_2$, and $\kappa_6/\kappa_2$ calculated using Monte Carlo sampling within the SAM.
The calculations incorporate the thermal smearing and resonance decays.
The results reveal large differences between net proton and net baryon cumulants ratios.
Net proton cumulant ratios are considerably closer to the Skellam baseline of unity.
This can be understood in the following way.
By taking only a subset of baryons -- the protons -- one dilutes the total signal due to baryon correlations.
This leads to a smaller deviation of cumulants from Poisson statistics -- the limiting case of vanishing correlations.

The large difference between net proton and net baryon cumulants clearly indicates that direct comparison between the two is not justified.
It is interesting that net proton cumulant ratios cross the grand-canonical value of the corresponding net baryon ratios in the grand-canonical limit~(horizonal lines in Fig.~\ref{fig:netbaryonproton}) for a sufficiently large acceptance.
This, for instance, takes place at $\Delta Y_{\rm acc} \simeq 1.4$ for $\kappa_2/\kappa_2^{\rm Skellam}$ while for $\kappa_4/\kappa_2$ the crossing is at $\Delta Y_{\rm acc} \simeq 2.5$.
The crossings take place due to suppression of net proton cumulants from baryon number conservation.
This accidental coincidence between net proton and grand-canonical net baryon cumulant ratios may be 
of relevance for the recent analysis of STAR data by the HotQCD collaboration in Ref.~\cite{Bazavov:2020bjn}.
There, the net baryon lattice QCD susceptibilities were directly compared to the measured net proton cumulants and an agreement, within large error bars, was reported.

We explore also, if the method of Kitazawa and Asakawa~\cite{Kitazawa:2011wh,Kitazawa:2012at} can be used to relate net proton and net baryon cumulants in the EV-HRG model.
To do that, we calculate net proton cumulants in an alternative way, namely by registering each baryon within the rapidity acceptance with a Bernoulli probability $q = \mean{p}/\mean{B}$. 
For an EV-HRG model at $T = 160$~MeV that we use one has $q\simeq 0.33$.
The net proton cumulants computed in this way are shown in Fig.~\ref{fig:netbaryonproton} by open blue diamonds.
They agree with the actual net proton cumulant ratios shown by blue circles.
This confirms that cumulants of net baryon distribution can be recovered from factorial moments of net proton distribution via a binomial unfolding with probability $q$.
The value of $q$ in experiment can be calculated from the measured mean multiplicities of the various baryon species.
The neutron yield, which is not measured, can be reconstructed from proton yields using the isospin symmetry.

\subsection{Comparison to ALICE data}

The results we have discussed so far correspond to fluctuations of baryons and protons in acceptances integrated over all transverse momenta.
This has not yet been achieved experimentally.
Instead, the ALICE collaboration has published measurements of the variance of net-proton distribution in Pb-Pb collisions at $\sqrt{s_{\rm NN}} = 2.76$~TeV in an acceptance in a 3-momentum range $0.6 < p < 1.5$~GeV/$c$ and longitudinal pseudorapidity  $|\eta| < 0.8$~\cite{Acharya:2019izy}.

\begin{figure}[t]
  \centering
\includegraphics[width=.49\textwidth]{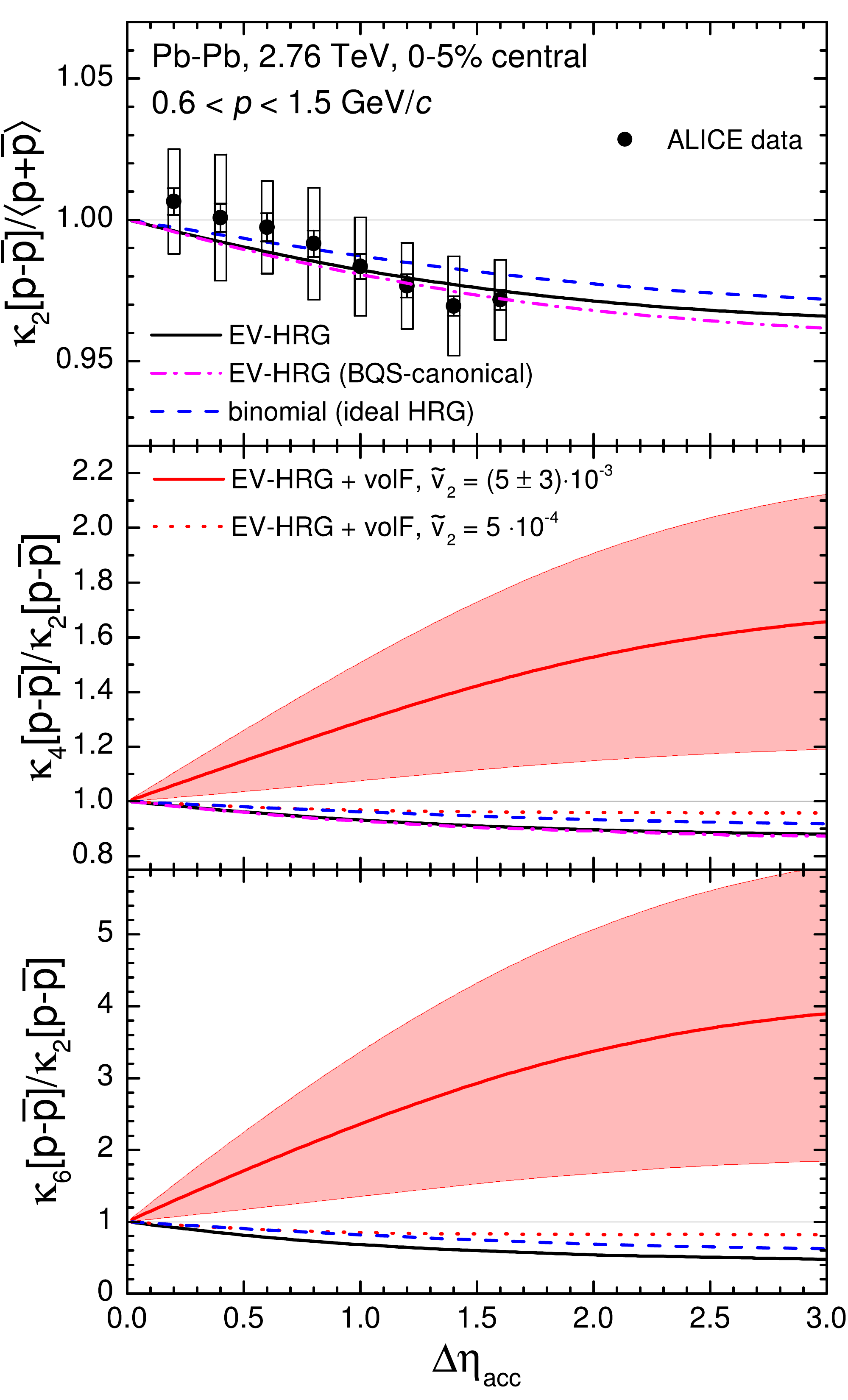}
  \caption{
   Pseudorapidity acceptance dependence of net proton cumulant ratios $\kappa_2/\kappa_2^{\rm Skellam}$ (top), $\kappa_4/\kappa_2$ (middle), and $\kappa_6/\kappa_2$ (bottom) in 0-5\% central Pb-Pb collisions at the LHC. 
   Calculations in an EV-HRG model with global baryon conservation while the dashed blue lines correspond to the binomial acceptance baseline.
   The red lines correspond to including the effect of volume fluctuations into the EV-HRG model, the band corresponds to the uncertainty in the variance of volume fluctuations~(see text).
   The dash-dotted magenta lines correspond to EV-HRG with additional conservation of electric charge and strangeness.
   The symbols depict the experimental data of the ALICE collaboration~\cite{Acharya:2019izy}.
  }
  \label{fig:netprotonsALICE}
\end{figure}

The top panel of Fig.~\ref{fig:netprotonsALICE} depicts the comparison between the data~(symbols) and the EV-HRG model with exact baryon number conservation~(black line) for the ratio $\kappa_2/\mean{p+\bar{p}}$ of net protons.
The data are described by the model within errors.
However, the data are described similarly well by the ideal HRG model, where this ratio is given by the binomial baseline, $(\kappa_2/\mean{p+\bar{p}})^{\rm bino} = 1 - \alpha_p$~\cite{Braun-Munzinger:2016yjz}. Here $\alpha_p = \mean{N_p}^{\rm acc} / \mean{N_B}^{4\pi}$ where $\mean{N_p}^{\rm acc}$ is the mean number of protons in the acceptance and $\mean{N_B}^{4\pi}$ is the mean number of baryons in the full space.
This implies that measurements in these acceptance windows are not very sensitive to the equation of state. The deviations from the Skellam baseline are overshadowed by the global baryon conservation.
The effect of repulsive interactions in the EV-HRG model is to slightly reduce the ratio further away from the Skellam limit.
This is in contrast to the baryon and proton cumulants in $p_T$-integrated acceptances that we have shown in Figs.~\ref{fig:netbaryonLHC} and~\ref{fig:netbaryonproton},
where the effect of interactions for one unit of rapidity is already sizable.
The reason is due to cuts in the transverse momentum coverage.
While the presence of radial flow does induce a level of correlation between the transverse momenta and coordinates of particles, this correlation is not as strong as in the longitudinal direction given by the Bjorken flow.
The $p_T$-cuts, therefore, lead to a Poissonization of the grand-canonical fluctuations, making it challenging to extract the grand-canonical susceptibilities.
This underlines the importance of expanding the acceptance for fluctuation measurements in the future runs at the LHC in order for them to be sensitive to the equation of state.

We also explore the effect of exact conservation of electric charge and strangeness of net proton fluctuations.
As shown in Ref.~\cite{Vovchenko:2020gne}, a moderate effect of these extra conservation laws on net proton cumulants is expected.
To evaluate the effect, we sample the grand-canonical multiplicities of all hadrons and resonances, including mesons, in the same fashion as before, but reject, in addition to baryon number conservation, all events which do not satisfy the exact conservation of global electric charge, $Q = 0$, and strangeness, $S = 0$.
These two additional rejection steps slow down the event generator procedure considerably.
Therefore, we generate a smaller number of events in the $BQS$-canonical ensemble, equaling to about $3\cdot 10^7$ events.
For this reason we restrict the analysis within the $BQS$-canonical ensemble to the second and fourth order cumulants.
The $\kappa_2/\mean{p+\bar{p}}$ ratio from the $BQS$-canonical EV-HRG model is depicted by a dash-dotted magenta line in the top panel of Fig.~\ref{fig:netprotonsALICE}.
The exact electric charge and strangeness conservation leads to a further reduction of $\kappa_2/\mean{p+\bar{p}}$ by a moderate amount.
This effect is consistent with results in reported in Ref.~\cite{Vovchenko:2020gne} using the SAM for multiple conserved charges.

The pseudorapidity dependencies of kurtosis and hyperkurtosis of net proton fluctuations within the same ALICE acceptance are depicted in the middle and bottom panels of Fig.~\ref{fig:netprotonsALICE}, respectively.
Similar to the variance, these show a suppression with respect to the Skellam baseline, mainly due to the baryon number conservation.
It is notable that the hyperkurtosis never reaches a negative value within the ALICE acceptance.
Again, this is a reflection of a limited $p_T$ coverage of the acceptance as well as of measuring only a subset of all baryons.

\subsection{Volume fluctuations}
\label{sec:volfluct}

We would like to discuss another issue which may affect fluctuation measurements in heavy-ion collisions, namely  fluctuations of the system volume.
The volume fluctuations do not affect the behavior of the mean quantities, but they do modify the fluctuations.
This effect has been studied in several works in the literature~\cite{Gorenstein:2011vq,Skokov:2012ds,Braun-Munzinger:2016yjz}.
Here we follow Ref.~\cite{Skokov:2012ds} to estimate the effect of volume fluctuation on our results.

We assume that, in the absence of volume fluctuations, all the cumulants  obey linear scaling with the volume, $\kappa_n \propto V$.
Let us denote by $\tilde{\kappa}_n$ the cumulants which include the effect of volume fluctuations.
They read~\cite{Skokov:2012ds}
\eq{\label{eq:volfluct}
\tilde{\kappa}_n = \sum_{l=1}^{n} V_l \, B_{n,l} (\kappa_1/V, \kappa_2/V, \ldots, \kappa_{n-l+1}/V)~.
}
Here $V_l$ is the $l$th cumulant of the system volume distribution and $B_{n,l}$ are Bell polynomials.

Let us now take into account that all odd-order cumulants of net-particle distribution at the LHC vanish, $\kappa_{2n-1} = 0$.
In this case the odd-order order cumulants with volume fluctuations do vanish as well, $\tilde{\kappa}_{2n-1}  = 0$.
The even order cumulants up to $n = 6$ read
\eq{\label{eq:volflucteven}
\tilde{\kappa}_2 & = \kappa_2, \\
\tilde{\kappa}_4 & = \kappa_4 + 3 \kappa_2^2 \,
\tilde{v}_2, \\
\tilde{\kappa}_6 & = \kappa_6 + 15 \kappa_2 \, \kappa_4 \, \tilde{v}_2 + 15 \kappa_2^3 \, \tilde{v}_3~.
}
Here $\tilde{v}_i = V_i /\mean{V}^i$ are the scaled volume cumulants.
The variance of a net-particle distribution at the LHC is not influenced by volume fluctuations, as pointed out before in Refs.~\cite{Skokov:2012ds,Braun-Munzinger:2016yjz}.
However, the volume fluctuations do influence the higher-order cumulants.

The cumulant ratios read
\eq{\label{eq:volratios}
\frac{\tilde{\kappa}_2}{\mean{p+\bar{p}}} & = \frac{\kappa_2}{\mean{p+\bar{p}}}~,\\
\label{eq:volratiokurt}
\frac{\tilde{\kappa}_4}{\tilde{\kappa}_2} & =
\frac{\kappa_4}{\kappa_2} + 3 \kappa_2 \, \tilde{v}_2, \\
\label{eq:volratioc6}
\frac{\tilde{\kappa}_6}{\tilde{\kappa}_2} & =
\frac{\kappa_6}{\kappa_2} + 15 \kappa_4 \, \tilde{v}_2 + 15 \kappa_2^2 \, \tilde{v}_3. 
}
A non-zero variance of the volume distribution influences the kurtosis and hyperkurtosis.
In addition, the hyperkurtosis may be affected by a non-zero skewness $\tilde{v}_3$ of the volume distribution.

The effect of volume fluctuations is determined by the values of the reduced cumulants $\tilde{v}_i$.
These are mainly determined by the collision geometry and the centrality selection.
To illustrate the effect of volume fluctuations we will consider net-proton fluctuations in the ALICE acceptance that we discussed in the previous subsection.
For simplicity, we shall neglect the skewness of volume fluctuations, $\tilde{v}_3 = 0$, which could have an influence on the hyperkurtosis, but not on the kurtosis.
To fix $\tilde{v}_2$ we make use of the ALICE measurement of the variance of proton number distribution~\cite{Acharya:2019izy}.
As the mean number of protons is non-vanishing even at the LHC energies, the variance $\tilde{\kappa}_2^p$ of proton number distribution is affected by the volume fluctuations, in contrast to net-proton variance which is unaffected.
Following Eq.~\eqref{eq:volfluct} the proton number scaled variance reads
\eq{\label{eq:kappa2prot}
\frac{\tilde{\kappa}_2^p}{\mean{p}} & = \frac{\kappa_2^p}{\mean{p}} + \mean{p} \, \tilde{v}_2
}

ALICE has measured $\tilde{\kappa}_2^p/\mean{p} = 1.07 \pm 0.06$ and $\mean{p} = 18.4 \pm 0.4$ in an acceptance $0.6 < p < 1.5$~GeV/$c$ and $|\eta| < 0.8$.
The EV-HRG model without volume fluctuations that we use, on the other hand, predicts $\kappa_2^p/\mean{p} = 0.98$.
Assuming that the difference between the model and the measurements can be attributed to volume fluctuations, one can use Eq.~\eqref{eq:kappa2prot} to extract the value of $\tilde{v}_2$ which describes the data:
\eq{\label{eq:v2}
\tilde{v}_2 = 0.005 \pm 0.003.
}

The pseudorapidity dependence of the kurtosis and hyperkurtosis of the net proton distribution in 0-5\% central Pb-Pb collisions at the LHC in the EV-HRG model with baryon number conservation and volume fluctuations in depicted in Fig.~\ref{fig:netprotonsALICE} by the red lines with bands.
The bands correspond to the error propagation of the variance of the volume distribution in Eq.~\eqref{eq:v2}.
The volume fluctuations have a large effect on higher-order fluctuations, both the kurtosis and hyperkurtosis exceed unity in all acceptances considered, in contrast to calculations without volume fluctuations where they lie below unity.
It seems, therefore, that a significant reduction of volume fluctuations will be required in the future experimental measurements to be able to reliably control this effect.
As an illustration, the dotted red lines in Fig.~\ref{fig:netprotonsALICE} depict the cumulant ratios when the variance of volume fluctuation is decreased by an order of magnitude, i.e. $\tilde{v}_2 = 5\cdot 10^{-4}$.
In this case, the results are considerably closer to the cumulant ratios without volume fluctuations, and it should be possible to reliably extract these ratios by fitting the data via Eqs.~\eqref{eq:volratios}-\eqref{eq:volratioc6}.
This type of analysis has been performed by the HADES collaboration in Ref.~\cite{Adamczewski-Musch:2020slf}, where the next-to-leading order volume dependence of the cumalants was additionally considered.
The centrality bin width correction~\cite{Luo:2013bmi} is another possible remedy, which has been applied for net proton measurements by the STAR collaboration~\cite{Adam:2020unf}.

\subsection{Net-$\Lambda$, net-kaon, and net-pion fluctuations}
\label{sec:piKL}

Net proton cumulants are not the only fluctuation measurement performed by the ALICE collaboration.
Fluctuations of net numbers of $\Lambda$'s, kaons and pions are also being performed, and preliminary results were reported in Refs.~\cite{Ohlson:2017wxu,Ohlson:2019erm,Arslandok:2020mda}.
Here we would like to discuss the behavior of these quantities within our approach.
The main purpose here is to illustrate how the different effects like resonance decays and exact conservation of various conserved charges influence the observables semi-quantitatively. 
Where available, we do confront our predictions with the preliminary data as well.
Our analysis here is restricted to the second cumulants normalized by the Skellam baselines, which at the LHC energies are free of the influence of volume fluctuations.

To perform the analysis we sample the full EV-HRG model, including both the (anti)baryons and mesons, using the same parameters as above.
The cumulants are calculated after all strong and electromagnetic decays, in the ALICE acceptance, $0.6 < p < 1.5$~GeV/$c$ and a pseudorapidity acceptance $|\eta| < 0.5 \, \Delta \eta_{\rm acc}$, where $\Delta \eta_{\rm acc}$ is varied up to a value of 3 units.
We consider three configurations for the treatment of global conservation laws: 
(i) global conservation laws are neglected~(grand-canonical);
(ii) exact conservation of baryon number is enforced~($B$-canonical);
(iii) exact conservation of baryon number, electric charge, and strangeness is enforced~($BQS$-canonical).
Comparing the results between the three cases allows us to distinguish the roles of different conservation laws.

\begin{figure}[t]
  \centering
\includegraphics[width=.49\textwidth]{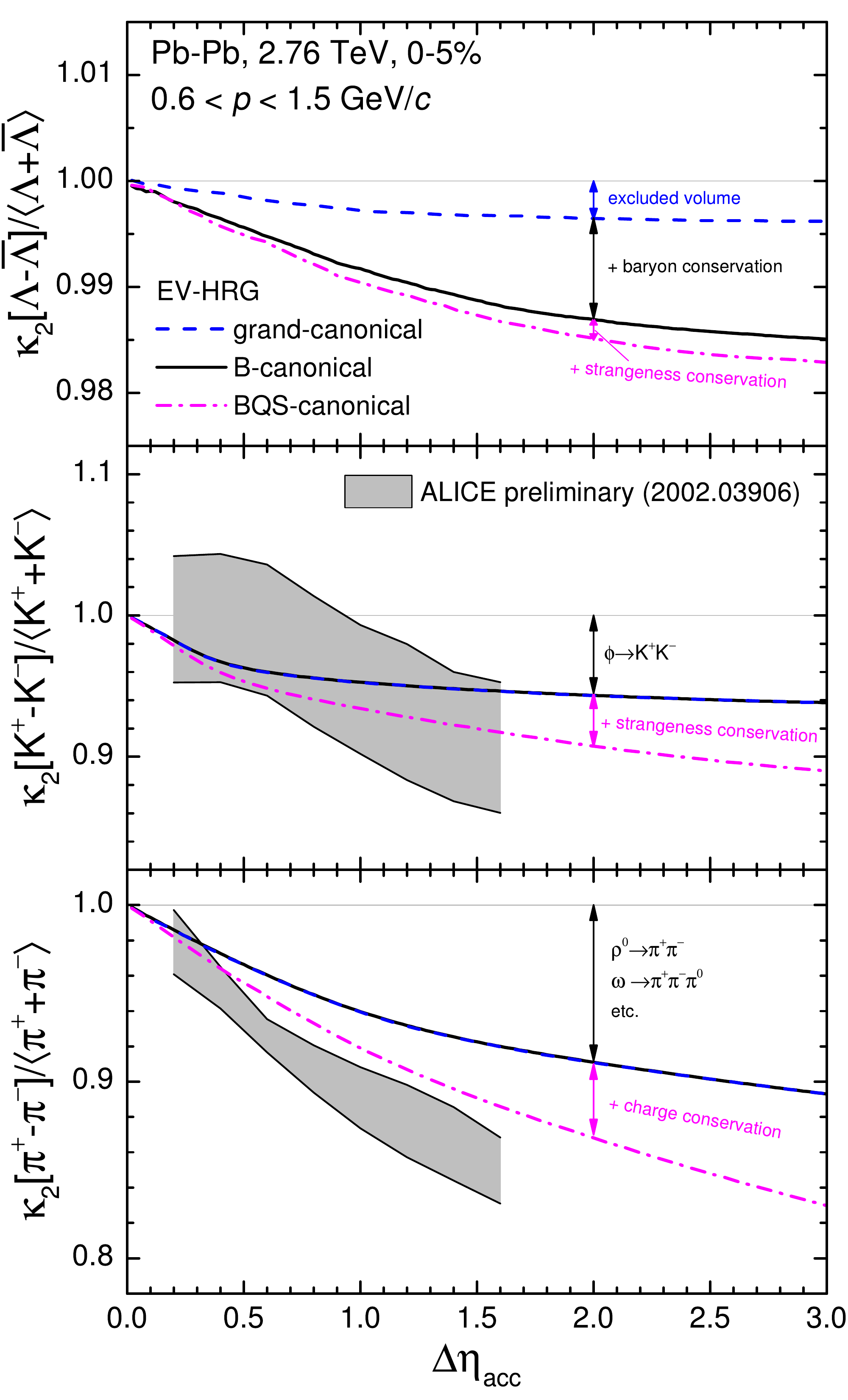}
  \caption{
   Pseudorapidity acceptance dependence of variance-over-Skellam ratios for net $\Lambda$~(top), net kaon~(middle), and net pion~(bottom) numbers in 0-5\% central Pb-Pb collisions at the LHC. 
   The lines depict calculations within the subensemble sampling of the EV-HRG model, without global conservation laws~(dashed blue lines), with global baryon conservation~(solid black lines), and with global conservation of baryon number, electric charge, and strangeness~(dash-dotted magenta lines).
   The bands represent the preliminary data of the ALICE collaboration~\cite{Ohlson:2017wxu,Arslandok:2020mda}.
  }
  \label{fig:netothersALICE}
\end{figure}

Let us start with the net-$\Lambda$ fluctuations.
The results are depicted in the top panel of Fig.~\ref{fig:netothersALICE}.
The $\ktwonet{\Lambda}$ ratio shows a mild suppression relative to unity as the pseudorapidity acceptance $\Delta \eta_{\rm acc}$ is increased.
A small suppression exists already in the grand-canonical limit~(dashed blue line), which is attributed to the presence of repulsive baryon interactions modeled by the excluded volume.
A larger effect is observed when the global baryon number conservation is incorporated~(solid black line).
An additional suppression from exact strangeness conservation on top of baryon conservation is also observed~(magenta line), although this effect is rather small.
This smallness is attributed to the fact that at LHC energies the dominant part of all strange quarks is carried by kaons, with $\Lambda$'s forming only a small fraction of all strange particles.

The net kaon fluctuations are interesting because they are affected by a decay $\phi \to K^+ K^-$ of the $\phi$ meson.
Our calculations, as well as experimental data~\cite{Abelev:2014uua}, suggest that about 6\% of final state $K^+$ and $K^-$ mesons come from this decay channel.
The decay generates a correlation between the numbers of $K^+$ and $K^-$.
If both decay products fall into a measurement acceptance, this gives no contribution to the variance $\kappa_2[K^+ - K^-]$ as the net number of kaons is unchanged.
However, the total number of charged kaons, $\mean{K^+ + K^-}$, increases by two.
For this reason one can expect the ratio $\kappa_2[K^+ - K^-]/\mean{K^+ + K^-}$ to be below unity due to resonance decays alone, even in the absence of global conservation laws.
This is indeed observed in our Monte Carlo simulations depicted in the middle panel of Fig.~\ref{fig:netothersALICE}: the ratio $\kappa_2[K^+ - K^-]/\mean{K^+ + K^-}$ is visibly below unity in the grand-canonical calculation which we attribute to the $\phi \to K^+ K^-$ decay.
The net kaon fluctuations are virtually unaffected by the exact baryon number conservation~(black line).
This is expected because mesons do not interact with the baryons in the EV-HRG model, hence the baryon number conservation does not have an influence on meson distribution, except for small feeddown contributions from baryonic resonances.

The kaons are affected by strangeness and, to a lesser extent, electric charge conservation.
The $BQS$-canonical calculation is depicted by the dash-dotted magenta line, showing a further suppression of the variance-over-Skellam ratio when strangeness and electric charge conservation is implemented.
The resulting $\Delta \eta_{\rm acc}$ dependence of net kaon fluctuations agrees with the preliminary data of the ALICE collaboration~\cite{Ohlson:2017wxu,Arslandok:2020mda}, shown in Fig.~\ref{fig:netothersALICE} by the gray bands, although the experimental uncertainties are quite large.

The behavior of net-pion fluctuations~(the bottom panel in Fig.~\ref{fig:netothersALICE}) is qualitatively similar to net kaons.
The pion fluctuations are affected more strongly by resonance decays than kaons.
Several resonances give a notable contribution.
The most notable ones are decays $\rho^0 \to \pi^+ \pi^-$, $\omega \to \pi^+ \pi^- \pi^0$, $\eta \to \pi^+ \pi^- \pi^0$, all leading to a sizable suppression of the ratio $\kappa_2[\pi^+ - \pi^-]/\mean{\pi^+ + \pi^-}$ relative to unity already in the grand-canonical limit~(the blue line).
Baryon conservation has a negligible influence on net pion fluctuations, similar to net kaon fluctuations.
Net pion fluctuations, however, are notably suppressed by the exact conservation of electric charge, see the dash-dotted magenta line.
This should not come as a big surprise, as the charged pions constitute the majority of all charged particles at the LHC, hence the sizable effect of charge conservation on pion fluctuations.

The preliminary data of the ALICE collaboration on net pion variance-over-Skellam ratio lies somewhat below our $BQS$-canonical model prediction, the deviations are roughly on a two-sigma level.
It should be cautioned that our predictions for net pion fluctuations should be regarded as semi-quantitative, for several reasons.
For instance, we use the blast-wave model parametrization from Ref.~\cite{Mazeliauskas:2019ifr} which underestimates significantly the number of soft pions, $p_T \lesssim 500$~MeV/$c$.
Also, we neglect the effect of Bose statistics, which is non-negligible for the primordial pions at the chemical freeze-out.
We also neglect additional effect due to rescattering in the hadronic phase.
It is known, however, that the number of $\rho^0$ resonances reconstructed in central Pb-Pb experimentally is about 20-25\% lower than predicted by the HRG model at the chemical freeze-out~\cite{Acharya:2018qnp,Motornenko:2019jha}.
This indicates additional dynamics in the hadronic phase involving $\rho^0$ resonances and their decay products, which may change the effect of $\rho^0$ decays on $\kappa_2[\pi^+ - \pi^-]/\mean{\pi^+ + \pi^-}$.
It is true however, that both the Bose statistics as well hadronic rescattering are expected to worsen the agreement with the data rather than improve it.
The Bose statistics leads to an enhancement of pion fluctuations~\cite{Begun:2008hq}, whereas the hadronic rescattering will dilute the correlations between pions from resonance decays in given acceptance, both effects thus leading to an increase of $\kappa_2[\pi^+ - \pi^-]/\mean{\pi^+ + \pi^-}$.
Nevertheless, our analysis is sufficient to indicate that net pion fluctuations are affected sizably by both the resonance decays as well as exact global conservation of electric charge.
Both these mechanisms should thus be taken into account in interpretations of experimental data.

\subsection{Dynamical net-charge fluctuations}

We would like to conclude our analysis of experimental data with the variance of the net-charge distribution. 
The corresponding measurements have been performed by the ALICE collaboration and published in Ref.~\cite{Abelev:2012pv}.
There, the measurements were focused on a quantity $\nu_{(+,-,\rm dyn)}$, defined as
\eq{
\nu_{(+,-,\rm dyn)} & = \frac{\mean{N_+ (N_+ - 1)}}{\mean{N_+}^2} + \frac{\mean{N_- (N_- - 1)}}{\mean{N_-}^2} \nonumber \\
& \qquad -2 \frac{\mean{N_+ N_-}}{\mean{N_+}\mean{N_-}}~.
}
Here $N_{+(-)}$ is the number of positively~(negatively) charged particles in the final state for a given acceptance.
In the limit $\mean{N_+} = \mean{N_-}$, which to a large precision holds at the LHC, $\nu_{(+,-,\rm dyn)}$ simplifies to
\eq{
\nu_{(+,-,\rm dyn)} = 4 \frac{\mean{\delta Q^2}}{\mean{N_{\rm ch}}^2} - \frac{4}{\mean{N_{\rm ch}}}~, \qquad \mean{N_+} = \mean{N_-}~.
}
Here $Q \equiv N_+ - N_-$ is the net charge and $\mean{N_{\rm ch}} \equiv \mean{N_+} + \mean{N_-}$ is the charged multiplicity.
$\nu_{(+,-,\rm dyn)}$ is thus closely related to the so-called $D$-measure:
\eq{\label{eq:D}
D & = \mean{N_{\rm ch}} \nu_{(+,-,\rm dyn)} + 4 \nonumber \\
 & = 4 \frac{\mean{\delta Q^2}}{\mean{N_{\rm ch}}}~.
}

The $D$-measure was introduced in Ref.~\cite{Jeon:2000wg} as a probe that discriminates the charge-carrier degrees of freedom in the medium. 
In the quark-gluon plasma~(QGP), where the quarks carry fractional charges, one has $D \sim 1-1.5$~\cite{Jeon:2000wg} in thermal equilibrium.
For a gas of hadrons and resonances, on the other hand, the baseline value is considerably larger, $D \sim 3-4$~\cite{Bleicher:2000ek}.

A direct comparison of the baselines with experimental measurements of net-charge fluctuations is complicated by several additional effects, including volume fluctuations, exact charge conservation, and acceptance cuts.
The situation at the LHC is favorable with regard to the volume fluctuations: as the $D$-measure~\eqref{eq:D} is defined by the variance of net charge fluctuations, it is unaffected by volume fluctuations due to an equal average numbers of positively and negatively charged particles, as discussed in Sec.~\ref{sec:volfluct}.
To account for the exact charge conservation, different corrections to Eq.~\eqref{eq:D} have been suggested in the literature.
Ref.~\cite{Pruneau:2002yf} advocated an additive correction:
\eq{\label{eq:Dp}
D' = D + 4 \, \alpha_{\rm ch}, \qquad \alpha_{\rm ch} = \frac{\mean{N_{\rm ch}}}{\mean{N_{\rm ch}^{4\pi}}}.
}
Here $\mean{N_{\rm ch}^{4\pi}}$ is the mean charged multiplicity in the full space.
Ref.~\cite{Bleicher:2000ek}, on the other hand, suggested a multiplicative correction:
\eq{\label{eq:Dpp}
D'' = \frac{D}{C_\mu \, (1 - \alpha_{\rm ch})}.
}
Here $1 - \alpha_{\rm ch}$ is the charge conservation correction factor while
$C_\mu = \mean{N_+}^2 / \mean{N_-}^2$~($=1$ at the LHC) additionally corrects for the effects of finite net charge.

The ALICE measurements in Ref.~\cite{Abelev:2012pv} include charge conservation corrections and incorporate the differences between $D'$ and $D''$ as a contribution to the systematic error.
Here we analyze the behavior of the $D$-measure within our Monte Carlo sampling of the EV-HRG model at LHC conditions.

\begin{figure}[t]
  \centering
  \includegraphics[width=.49\textwidth]{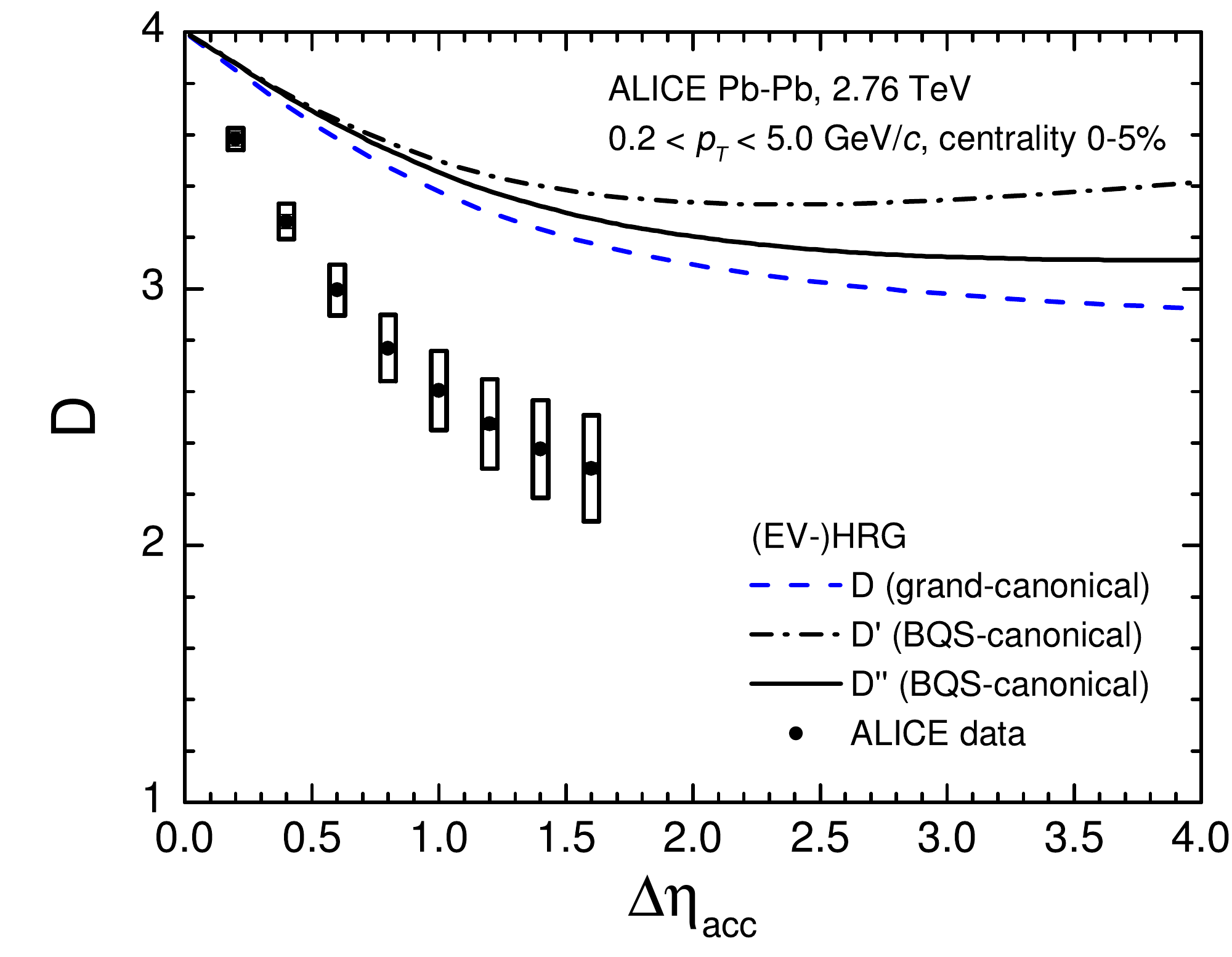}
  \caption{
   Pseudorapidity acceptance dependence of the $D$-measure of the dynamical net-charge fluctuations in 0-5\% central Pb-Pb collisions at the LHC. 
   The dashed blue line depicts calculations within the Monte Carlo sampling of the EV-HRG model without global conservation laws~(the grand-canonical ensemble).
   The black lines correspond to EV-HRG model calculations with exact conservation of baryon number, electric charge, and strangeness, where the $D$-measure is corrected for exact charge conservation in an additive~(dash-dotted, following Ref.~\cite{Pruneau:2002yf}) or multiplicative~(solid, following Ref.~\cite{Bleicher:2000ek}) ways.
   The symbols depict the experimental data of the ALICE collaboration~\cite{Abelev:2012pv} corrected for charge conservation.
  }
  \label{fig:nudyn}
\end{figure}

Figure~\ref{fig:nudyn} depicts the pseudorapidity acceptance dependence of the $D$-measure of the dynamical net-charge fluctuations in 0-5\% central Pb-Pb collisions at the LHC calculated in the EV-HRG within various statistical ensembles.
A transverse momentum cut $0.2 < p_T < 5.0$~GeV/$c$ is applied.
This is the same $p_T$ cut as in the ALICE measurement.
The dashed blue line depicts the behavior of the $D$-measure in the grand-canonical version of the EV-HRG model.
As the grand-canonical calculation neglects the exact charge conservation, we calculate the $D$-measure in this case directly using Eq.~\eqref{eq:D}, without applying any of the charge conservation corrections.
The resulting $D$-measure is a decreasing function of $\Delta \eta_{\rm acc}$ that saturates at a value of around $D \sim 2.8$ in the limit $\Delta \eta_{\rm acc} \to \infty$.
The suppression of $D$ relative to the Poisson statistics baseline of $D = 4$ is attributed to decays of neutral resonances into a pair of charged particles, like $\rho^0 \to \pi^+ \pi^-$.
Here the discussion of resonance decays affecting net-pion fluctuations in Sec.~\ref{sec:piKL} straightforwardly applies.
We note that the influence of the excluded-volume effects in the baryon sector is virtually negligible, as the majority of charged particles at the LHC are mesons.
Therefore, the results shown in Fig.~\ref{fig:nudyn} for the EV-HRG also apply to the standard ideal HRG model.

Calculations incorporating exact conservation of various conserved charges reveal that net-charge fluctuations are affected by exact conservation of the electric charge, while the additional influence of baryon number and strangeness conservation is observed to be negligible.
This observation is consistent with the results of Ref.~\cite{Vovchenko:2020gne}, where it was shown that the variance of a conserved charge distribution is only affected by exact conservation of that charge, but not of any other conserved charge.
The black lines in Fig.~\ref{fig:nudyn} show the results of the $BQS$-canonical EV-HRG model calculation where we apply the charge conservation correction in accordance with Eq.~\eqref{eq:Dp}~[$D'$, dash-dotted line] or~\eqref{eq:Dpp}~[$D''$, solid line].
This is the same procedure that was performed by the ALICE collaboration in Ref.~\cite{Abelev:2012pv} to correct for global charge conservation.
If these corrections were exact, one would expect to reproduce the grand-canonical result shown by the dashed blue line.
Instead, we observe that both the $D'$ and $D''$ appear to overestimate the charge conservation correction, especially $D'$ at large $\Delta \eta_{\rm acc}$.
The $D''$ correction does perform better than $D'$ and stays close to the grand-canonical result for $\Delta \eta_{\rm acc} \lesssim 4$.

The experimental data of the ALICE collaboration are shown by the symbols with error bars in Fig.~\ref{fig:nudyn}.
The data points lie considerably lower than model predictions.
In particular, the slope of the curve at small $\Delta \eta_{\rm acc}$ is much steeper in the data than in the model.
This result is in line with the tensions of the HRG model with the preliminary data for net-pion fluctuations discussed in Sec.~\ref{sec:piKL}.
The visibly stronger effect obtained for the $D$-measure can be attributed to a significantly larger transverse momentum coverage for the net-charge fluctuations relative to those for net pions.
As discussed in Sec.~\ref{sec:piKL}, the effects that we neglected in our calculations, such as the Bose-Einstein statistics for pions or hadronic rescattering, would be expected to enhance the $D$-measure and thus even further worsen the disagreement with the data.
At this point we do not see a conceivable mechanism to explain the ALICE data within a purely hadronic description. 
The measurement, therefore, points to the suppression of net-charge fluctuations in central heavy-ion collisions at the LHC relative to the hadronic scenario.
One tantalizing possibility here is the QGP formation, where a suppression of the $D$-measure is expected~\cite{Jeon:2000wg}.
We hope that future measurements and analyses will shed more light on whether the observation of a suppressed $D$-measure constitutes a QGP signature.

\section{Discussion and summary}
\label{sec:summary}

In this work we introduced a subsensemble sampler -- a novel particlization routine for heavy-ion collisions which preserves the thermal fluctuations and correlations in an interacting hadron resonance gas on a local level. 
It also takes into account global conservation laws on an event-by-event basis.
The key of the procedure lies in partitioning the particlization hypersurface into locally grand-canonical subvolumes.
In each subvolume, the hadron numbers are sampled from the grand-canonical multiplicity distribution, while their momenta follow from a thermal distribution imposed on a collective velocity profile.
The global conservation laws are enforced via a subsequent rejection sampling step.
The procedure allows to evaluate event-by-event fluctuations of various particle numbers in a momentum space acceptance, as appropriate for experiment, within a fluid dynamical picture of a local thermodynamic equilibrium at particlization.

The partition into subvolumes is not unique, the choice can be optimized for the applications on hand.
Certain restrictions do apply.
On the one hand, each subvolume $V_i$ has to be sufficiently large such that the cumulants of hadron multiplicity distribution are in the regime where they scale linearly with $V_i$.
On the other hand, the partition should be sufficiently fine grained, both relative to the acceptance where measurements are performed as well as to capture the coordinate space inhomogeneities in the  distribution of thermal parameters.
In the present work we considered the partition along the space-time rapidity axis~(Fig.~\ref{fig:partition}), which is appropriate to study the rapidity dependence of fluctuations integrated over the transverse momenta.
Other partition schemes can be considered in a more general case.

As a first application of our new particlization routine, we studied event-by-event fluctuations in Pb-Pb collisions at the LHC, with a focus on the rapidity acceptance dependence of cumulants of the net baryon distribution.
To that end, we utilized a hadron resonance gas model with excluded volume interactions in the baryonic sector, which matches well the available lattice QCD data on cumulants of net baryon distribution at a particlization temperature of $T = 160$~MeV.
We used a blast-wave flow velocity profile and neglected any dynamics in the hadronic phase except for strong and electromagnetic decays of resonances.

Our Monte Carlo simulations reveal how baryon interactions, global baryon conservation, thermal smearing, and resonance decays affect the behavior of cumulants in a momentum acceptance around midrapidity, as is appropriate for experimental measurements.
In the absence of thermal smearing and resonance decays, net baryon cumulants follow the analytic baseline established within a subensemble acceptance method~(SAM) in Ref.~\cite{Vovchenko:2020tsr}.
One can therefore use the SAM to correct experimental measurements for the effects of global baryon conservation.
However, for this to work the experimental acceptance needs to have a sufficiently large rapidity coverage, roughly $\Delta Y_{\rm acc} \gtrsim 1$, and capture the entire transverse momentum range.
The reason for that is the thermal smearing, which dilutes the signal for small acceptance and causes the cumulant ratios to approach the binomial distribution baseline~(see the red points in Fig.~\ref{fig:netbaryonLHC}).
We do observe that this effect is well described at LHC by a simplified analytic model which assumes the thermal smearing in kinematical rapidity to be Gaussian~(see Appendix).
The resulting expressions are somewhat more involved than the simple formulas of the pure SAM framework, but it is possible they can be used to subtract the effect of thermal smearing from the data in addition to global conservation.
The effect of an additional rapidity smearing of baryons due to decays of resonances is found to be negligible.

We find large differences between experimentally measurable net proton cumulants and the theoretically calculated net baryon cumulants.
The net proton cumulants generally lie much closer to the Skellam baseline than the net baryon cumulants.
This is a reflection of the fact that protons form a subset of all baryons. 
Measuring a subset, as opposed to the full set, dilutes the strength of correlations, which is reflected by the difference between net proton and net baryon cumulants in Fig.~\ref{fig:netbaryonproton}.
We do observe that net proton and net baryon cumulants are indeed related to each other by a binomial (un)folding, as advocated by Kitazawa and Asakawa~\cite{Kitazawa:2011wh,Kitazawa:2012at}.
This result, however, does \emph{not} by any means imply that one can directly compare net proton cumulant ratios with the net baryon ones.
Such a comparison is not only unjustified, but is likely to lead to misleading interpretations and conclusions.
For meaningful comparisons one has to reconstruct the net baryon cumulants from net proton ones through the binomial unfolding procedure described in~\cite{Kitazawa:2011wh,Kitazawa:2012at}. 
This has not yet been achieved in the present experiments although the procedure is doable and, in fact, straightforward, requiring the use of the factorial moments of (anti)proton distributions that are readily accessible in experiment.
On the other hand, the factorial moments of baryons and antibaryons are not directly accessible in lattice QCD, therefore, applying the method of Kitazawa and Asakawa to construct net proton cumulants from the lattice results on net baryon cumulants requires model assumptions.
This observation underscores the importance of measuring the factorial moments of (anti)proton distribution in addition to net proton cumulants, as only in this case one can reconstruct the net baryon cumulants and make the comparisons with various theoretical predictions meaningful.

We confronted the predictions of our event generator with the available experimental data of the ALICE collaboration on the variance of net proton, net pion, net kaon, and net charge distributions.
We find good agreement of our event generator with the net-proton data.
However, the data are described similarly well by the binomial distribution baseline that corresponds to an ideal hadron gas model with baryon number conservation.
In other words, the currently available measurements, performed in a 3-momentum and pseudorapidity acceptance, do not allow to distinguish the subtle effects associated with the QCD chiral crossover transition.
The variances of net-pion and net-kaon fluctuations are not sensitive to the interactions in the baryonic sector and global baryon conservation, but they are affected by resonance decays and exact conservation of electric charge and strangeness.
Our model describes the preliminary ALICE data on net kaon fluctuations within error bars.
The model also describes the trends seen in the pseudorapidity acceptance of net-pion fluctuations although the preliminary data are overestimated roughly on a two-sigma level.

The HRG model we employ does not describe the ALICE data on net-charge fluctuations. 
The experimental data on the $D$-measure is significantly below the model predictions~(Fig.~\ref{fig:nudyn}).
It seems doubtful that the measurement can be described within a purely hadronic description.
A suppression of the $D$-measure, on the other hand, is expected in quark-gluon plasma phase~\cite{Jeon:2000wg}. 
In fact, this has been the primary motivation for the corresponding measurements.
It remains to be seen whether the ALICE measurement is indeed a signal of QGP.

Measurements of higher-order cumulant ratios are affected by volume fluctuations.
We estimated the effect for 0-5\% central 2.76~TeV Pb-Pb collisions based on the available data of the ALICE collaboration on the first two proton number cumulants and the volume fluctuations formalism of Ref.~\cite{Skokov:2012ds}.
We found the effect to be quite large for the kurtosis and hyperkurtosis of net proton fluctuations in the ALICE acceptance, changing the qualitative nature of the pseudorapidity window dependence of these observables.
Therefore, removing the contribution of volume fluctuations will be essential for interpreting the experimental data, and our results indicate that the centrality selection should be optimized in the future LHC measurements of the higher-order net proton fluctuations.

The formalism developed in this work has many future applications.
One natural extension are the studies of fluctuations at lower collision energies probed by beam energy scan programmes at RHIC and SPS facilities~\cite{Bzdak:2019pkr}.
There, the effects of finite (baryo)chemical potentials, nonuniform rapidity distribution of thermal parameters, and absence of the longitudinal boost invariance will play an additional role~\cite{Biedron:2006vf,Becattini:2007ci}.
One can also consider a particlization hypersurface and flow velocity profile emerging from a full (3+1)-dimensional hydro simulation as opposed to the blast-wave model that we used here, which may additionally include viscous corrections.
It would also be of interest to analyze the effect of rescaterrings in the hadronic phase which would enhance the effect of momentum smearing and thus dilute the signal~\cite{Steinheimer:2016cir}.
This can be achieved by coupling the particlization to a subsequent hadronic  afterburner such as UrQMD or SMASH.


\begin{acknowledgments}

V.V. acknowledges the support through the
Feodor Lynen program of the Alexander von Humboldt
foundation.
This work received support through the U.S. Department of Energy, 
Office of Science, Office of Nuclear Physics, under contract number 
DE-AC02-05CH11231231 and within the framework of the
Beam Energy Scan Theory (BEST) Topical Collaboration.
The computational resources were provided by the Kronos computing cluster at GSI.

\end{acknowledgments}

\appendix

\section*{Appendix}

\subsection*{An analytic model to account for momentum smearing in net baryon cumulants}
\label{app:smearing}

Here we present a simplified analytic model to account for the effect of momentum smearing on the cumulants of net baryon distribution measured in a $p_T$-integrated acceptance.
The formalism here is applicable for interacting HRG models where correlations between numbers of baryons and antibaryons are absent in the grand-canonical ensemble.
This is the case, for instance, for the EV-HRG model that we use in this study.

The model consists of two steps: (i) the effect of momentum smearing is evaluated in the grand-canonical ensemble, i.e. neglecting the exact baryon number conservation; (ii) the SAM framework is applied to the result obtained in the first step to incorporate the exact baryon number conservation.

Let us start with the first part of the procedure.
Consider all baryons and antibaryons at particlization that have a longitudinal space-time coordinate $\eta_s$ within a narrow range $[\eta_s^0 - \Delta \eta_s/2, \eta_s^0 + \Delta \eta_s/2]$.
We assume that the physical volume corresponding to this range is large enough to capture all the physics associated with the correlation length, i.e. $dV/d\eta_s \, \Delta \eta_s \gg \xi^3$.
This means that, in the absence of exact baryon number conservation, the distribution of (net) baryons from this space-time rapidity range is independent from all other particles outside this range and is determined by the grand-canonical susceptibilities, namely
\eq{
\kappa_{n}^{B, \rm gce} (\eta_s^0) & = 
dV/d\eta_s \, \Delta \eta_s \, T^3 \, \chi_n^B,
\qquad |\eta_s -\eta_s^0| < \Delta \eta_s~.
}

In the absence of correlations between numbers of baryons and antibaryons in the grand-canonical ensemble that we assumed, the susceptibilities and cumulants are partitioned as follows:
\eq{
\chi_n^B &= \chi_n^{B^+} + (-1)^n \, \chi_n^{B^-}, \\
\kappa_{n}^{B,\rm gce}(\eta_s^0) & = \kappa_n^{B^+, \rm gce}(\eta_s^0) + (-1)^n \, \kappa_n^{B^-, \rm gce}(\eta_s^0).
}
Here $\chi_n^{B^+}$ and $\chi_n^{B^-}$ are the grand-canonical susceptibilities of baryon and antibaryon number, respectively.

Consider now the baryons which end up in a longitudinal rapidity acceptance $|Y| < \Delta Y_{\rm acc} / 2$.
Since the contributions from the different $\Delta \eta_s$ slices are independent, the resulting cumulants of the accepted particles are a sum of the contributions from the individual slices.
Therefore, let us calculate the contribution from a single slice $|\eta_s -\eta_s^0| < \Delta \eta_s$.
Let us denote by $p(\eta_s^0, \Delta Y_{\rm acc})$ the probability that a baryon with a space-time rapidity $\eta_s^0$ at particlization ends up in this acceptance.
This probability is determined by thermal smearing.
Assuming that all (anti)baryons at a given space-time rapidity $\eta_s^0$ end up in the kinematical acceptance independently from each other and approximating the probability $p(\eta_s^0, \Delta Y_{\rm acc})$ to be uniform in a range $\Delta \eta_s$, the cumulants of distribution of (anti)baryons in acceptance $|Y| < \Delta Y_{\rm acc} / 2$ that came from the space-time rapidity range $|\eta_s -\eta_s^0| < \Delta \eta_s$ are obtained by applying a binomial filter with the Bernoulli probability $p(\eta_s^0, \Delta Y_{\rm acc})$ to the space-time rapidity cumulants $\kappa_l^{B^{\pm}, \rm gce} (\eta_s^0)$:
\eq{
\kappa_n^{B^{\pm}, \rm gce}(\Delta Y_{\rm acc}, \eta_s^0) & = 
k_n^{\rm bino} [p(\eta_s^0, \Delta Y_{\rm acc}), \{ \kappa_l^{B^{\pm}, \rm gce} (\eta_s^0) \}].
}
Here $k_n^{\rm bino}[p,\{\kappa_l\}]$ is a $n$th order cumulant of particle number distribution obtained by applying the binomial filter with probability $p$ to a distribution described by a set of cumulants $\{\kappa_l\}$ with $l = 1 \ldots n$.
The cumulant generating function $C_{k}(t)$ for the cumulants $k_n$ after the binomial filter can be expressed in terms of the corresponding cumulant generating function $C_{\kappa}(t)$ for cumulants $\kappa$ before the filter~\cite{Kitazawa:2016awu,Savchuk:2019xfg}:
\eq{
C_{k^{\rm bino}}(t) = C_{\kappa}[\ln(1 - p + e^t p)]~.
}
The explicit result for the first four cumulants reads
\eq{
k_1^{\rm bino} & = p \, \kappa_1,\\
k_2^{\rm bino} & = p^2 \, \kappa_2 + p(1-p) \, \kappa_1,\\
k_3^{\rm bino} & = p^3 \, \kappa_3 + p(1-p)\left[3p \, \kappa_2 + (1-2p) \, \kappa_1 \right],\\
k_4^{\rm bino} & = p^4 \, \kappa_4 + p(1-p) 
\left\{ 
6p^2 \, \kappa_3
+ p (7-11p) \, \kappa_2
\right. \nonumber \\
& \quad \left. + [1-6p(1-p)] \, \kappa_1 \right\}~.
}

As already mentioned, the full result for cumulants $\kappa_n^{B^{\pm}, \rm gce}(\Delta Y_{\rm acc})$ of all (anti)baryons in the rapidity acceptance is obtained by summing the contributions from all $\eta_s$ slices. One obtaines
\eq{
\kappa_n^{B^{\pm}, \rm gce}(\Delta Y_{\rm acc}) & =
\int d \eta_s \, k_n^{\rm bino} \left[p(\eta_s, \Delta Y_{\rm acc}), \frac{d \kappa_n^{B^\pm, \rm gce}}{d \eta_s} \right],
}
where
\eq{
\frac{d \kappa_n^{B^\pm, \rm gce}}{d \eta_s} = \frac{dV}{d \eta_s} \, T^3 \, \chi_l^{B^{\pm}}~.
}
Note that $dV/d \eta_s$, $T$, and $\chi_l^{B^{\pm}}$ can all depend on $\eta_s$ in general case.
The net baryon cumulant is then simply
\eq{\label{eq:kappasmeargce}
\kappa_n^{B, \rm gce}(\Delta Y_{\rm acc}) = \kappa_n^{B^+, \rm gce}(\Delta Y_{\rm acc}) + (-1)^n \, \kappa_n^{B^-, \rm gce}(\Delta Y_{\rm acc})~.
}

How to evaluate the binomial probability $p(\eta_s, \Delta Y_{\rm acc})$?
We shall assume that the kinematical rapidity of each baryon is smeared around the space-time rapidity coordinate $\eta_s$ in accordance with a Gaussian distribution with a width $\sigma_y$.
The width can be estimated by analyzing the flow velocity and temperature profiles at the particlization hypersurface.
For the blast-wave model that we use at the LHC one has $\sigma_y \approx 0.3$.
The binomial probability reads
\eq{\label{eq:psmear}
p(\eta_s, \Delta Y_{\rm acc}) = 
\int\displaylimits_{-\Delta Y_{\rm acc}/2}^{\Delta Y_{\rm acc}/2} \,  dY \, \frac{\exp\left[-\frac{(Y - \eta_s)^2}{2 \sigma_y^2} \right]}{\sqrt{2\pi} \sigma_y}~.
}

Equations~\eqref{eq:kappasmeargce} and~\eqref{eq:psmear} allow to calculate the influence of thermal smearing in the grand-canonical ensemble.
In order to incorporate the exact conservation of baryon number we apply the SAM framework of Ref.~\cite{Vovchenko:2020tsr}.
The canonical ensemble cumulants that include both the effect of thermal smearing and global baryon conservation read
\eq{
& \kappa_1^{B, \rm ce} = \kappa_1^{B, \rm gce},\\
& \kappa_2^{B, \rm ce} = \beta \, \kappa_2^{B, \rm gce}, \\
& \kappa_3^{B, \rm ce} = \beta \, (1-2\alpha) \kappa_3^{B, \rm gce},\\
& \kappa_4^{B, \rm ce} = \beta (1-3\alpha \beta) \, \kappa_4^{B, \rm gce} - 3 \alpha \beta^2 \frac{(\kappa_3^{B, \rm gce})^2}{\kappa_2^{B, \rm gce}}, \\
& \kappa_5^{B, \rm ce}  = \beta \, (1-2\alpha) \left\{[1-2\beta\alpha]\kappa_5^{B, \rm gce}~ \right. \nonumber \\
& \quad \left. - 10 \alpha \beta \frac{\kappa_3^{B, \rm gce}  \kappa_4^{B, \rm gce}}{\kappa_2^{B, \rm gce}}\right\}, \\
& \kappa_6^{B, \rm ce}  = \beta \left[1-5\alpha \beta (1 - \alpha \beta ) \right]\kappa_6^{B, \rm gce} + 5 \, \alpha \, \beta^2 \nonumber \\
& \quad \times \left\{9 \alpha \beta \frac{(\kappa_3^{B, \rm gce})^2 \, \kappa_4^{B, \rm gce}}{(\kappa_2^{B, \rm gce})^2} - 3 \alpha \beta \frac{(\kappa_3^{B, \rm gce})^4}{(\kappa_2^{B, \rm gce})^3}  \right. \nonumber \\
& \quad \quad \left.  - 2 (1-2\alpha)^2 \frac{(\kappa_4^{B, \rm gce})^2 }{\kappa_2^{B, \rm gce}} - 3[1 - 3\beta \alpha] \frac{\kappa_3^{B, \rm gce} \, \kappa_5^{B, \rm gce}}{\kappa_2^{B, \rm gce}} \right\}.
}
Here $\alpha$ is the fraction of the total volume which is covered by the acceptance and $\beta \equiv 1 - \alpha$.
For the LHC energies that we study in this paper $\alpha = \Delta Y_{\rm acc} / (2 \eta_s^{\rm max})$.
It is also implied $\kappa_n^{B, \rm ce(gce)} \equiv \kappa_n^{B, \rm ce(gce)}(\Delta Y_{\rm acc})$ in the equations above.

We would like to emphasize again that the thermal smearing model here is based on the assumption that numbers of baryons and antibaryons are uncorrelated in the grand-canonical limit.
While this is the case for the EV-HRG model that we use in the present paper, this is not necessarily the case for other theories.
Modifying the smearing model to the general case should be possible, and will require the use of binomial filtering applied to factorial moments of baryon and antibaryon distributions, as discussed in~\cite{Bzdak:2012ab} in the context of acceptance corrections to net baryon and net proton cumulants.
It should also be possible to generalize the model to $p_T$-differential measurements and smearing based on thermal distributions superimposed on a realistic flow velocity profile and a 3-dimensional particlization hypersurface, as appropriate for the differential momentum distribution measurements in experiment.

\bibliography{SAMpler}


\end{document}